\documentclass[final]{elsarticle}

\pdfoutput=1

\usepackage{color}
\usepackage{times, amssymb, amsmath}
\usepackage{graphicx}
\usepackage{lineno}

\newcommand{\eps}{\varepsilon}

\newcommand{\beq}{\begin{equation}}
\newcommand{\eeq}{\end{equation}}

\newcommand{\dd}{\mbox{d}}

\begin{document}

\DeclareGraphicsExtensions{.jpg,.pdf,.png,.mps}


\begin{frontmatter}

\title{Particle identification}

\author{Christian~Lippmann}
\ead{C.Lippmann@gsi.de}
\address{GSI Helmholtzzentrum f\"{u}r Schwerionenforschung,
  Planckstra\ss e 1, 64291 Darmstadt, Germany}


\begin{abstract}

  Particle IDentification (PID) is fundamental to particle physics
  experiments. This paper reviews PID strategies and methods used by
  the large LHC experiments, which provide outstanding examples of the
  state-of-the-art. The first part focuses on the general design of these
  experiments with respect to PID and the technologies used. Three
  PID techniques are discussed in more detail: ionization
  measurements, time-of-flight measurements and Cherenkov
  imaging. Four examples of the implementation of these techniques at
  the LHC are given, together with selections of relevant examples
  from other experiments and short overviews on new
  developments. Finally, the Alpha Magnetic Spectrometer (AMS\,02)
  experiment is briefly described as an impressive example of a
  space-based experiment using a number of familiar PID techniques.

\end{abstract}

\begin{keyword}
  PID \sep particle identification \sep d$E$/d$x$ \sep ionization \sep
  TPC \sep Time Projection Chamber\sep RICH \sep Cherenkov radiation
  \sep Cherenkov ring imaging \sep time-of-flight \sep TOF \sep MRPC
  \sep separation power
\end{keyword}

\end{frontmatter}

\section{Introduction}
\label{sec:intro}

In addition to tracking and calorimetry, Particle IDentification (PID)
is a crucial aspect of most particle physics experiments. The
identification of stable particles is achieved either by analysing
the way they interact, or by determining their mass. The difference in
interaction is primarily used for lepton and photon identification. A
``traditional'' particle physics experiment already incorporates 
this method in its conceptual design, as it is discussed in
Sect.~\ref{sec:classic}. In order to unambiguously identify hadrons,
their charge and mass has to be determinded, which is achieved by
simultaneous measurements of momentum and velocity. This method is
discussed in Sect.~\ref{sec:chPID}.

The four large experiments at the CERN Large Hadron Collider
(LHC)~\cite{lhc} and their PID strategies are introduced briefly in
Sect.~\ref{sec:overview}. Three methods to determine the mass of a
charged particle and examples of their implementation at the LHC
experiments are discussed in Sects.~\ref{sec:dedx} (ionization),
\ref{sec:tof} (time-of-flight) and \ref{sec:rich} (Cherenkov radiation
imaging). A fourth important method, the detection of
transition radiation, is not discussed, since it is the topic of a 
separate paper in this issue~\cite{trds}. In Sect.~\ref{sec:nonacc} finally
the AMS\,02 experiment is briefly described as an example for the usage
of PID techniques from accelerator-based particle physics experiments
in space.

\subsection{PID by difference in interaction}
\label{sec:classic}

\begin{figure}[t]
  \begin{center}
    \includegraphics[width=8cm]{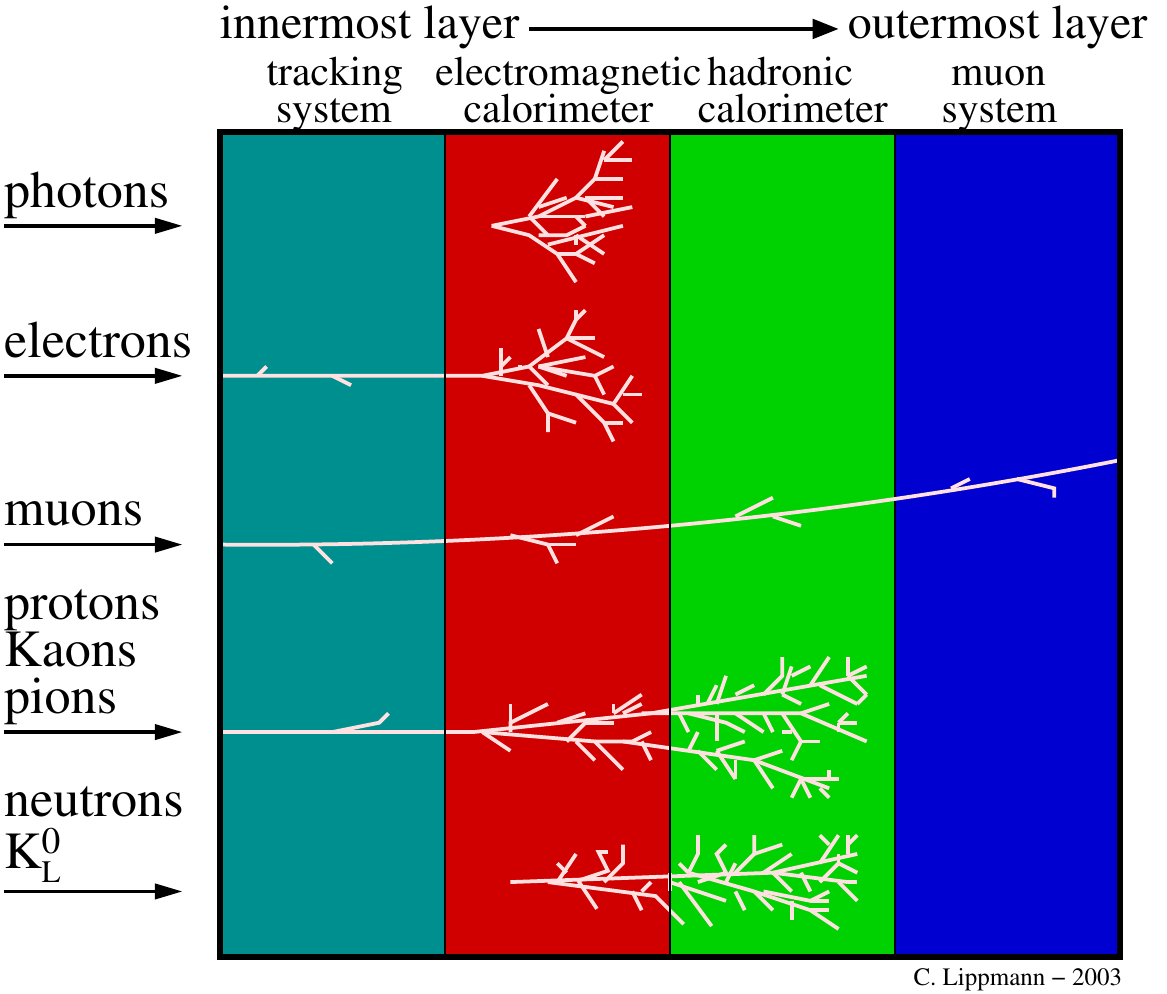}
    \caption{Components of a ``traditional'' particle physics
      experiment. Each particle type has its own signature in the
      detector. For example, if a particle is detected only in the
      electromagnetic calorimeter, it is fairly certain that it is a
      photon.}
    \label{fig:exp}
  \end{center}
\end{figure}

In a ``traditional'' particle physics experiment particles are
identified (electrons and muons, their antiparticles and photons),
or at least assigned to families (charged or neutral hadrons), by the
characteristic signatures they leave in the detector. The experiment
is divided into a few main components, as shown in Fig.~\ref{fig:exp},
where each component tests for a specific set of particle
properties. These components are stacked in layers and the particles
go through the layers sequentially from the collision point outwards:
first a tracking system, then an electromagnetic (EM) and a hadronic
calorimeter and a muon system. All layers are embedded in a magnetic
field in order to bend the tracks of charged particles for momentum
and charge sign determination.

\subsubsection{Tracking system}

The tracking system determines whether the particles are charged.
In conjunction with a magnetic field, it measures the sign of the
charge and the momentum of the particle. Photons may convert
into an electron-positron pair and can in that case be detected in the
tracking system. Moreover, charged kaon decays may be detected
in a high-resolution tracking system through their characteristic
``kink'' topology:~e.g.
K$^\pm \rightarrow \mu^\pm\nu_\mu$ (64\%) and 
K$^\pm \rightarrow \pi^\pm\pi^0$ (21\%).
The charged parent (kaon) decays into a neutral daughter (not detected)
and a charged daughter with the same sign. Therefore, the kaon
identification process reduces to the finding of kinks in the tracking
system. The kinematics of this kink topology allows to separate kaon
decays from the main source of background kinks coming from
charged pion decays~\cite{kink}.

\subsubsection{Calorimeters}

Calorimeters detect neutral particles, measure the energy of particles,
and determine whether they have electromagnetic or hadronic
interactions. EM and hadron calorimetry at the LHC is described in
detail in Refs.~\cite{ecal,hcal}. PID in a calorimeter is a destructive
measurement. All particles except muons and neutrinos deposit all
their energy in the calorimeter system by production of
electromagnetic or hadronic showers. The relative resolution with
which the deposited energy is measured is usually parametrised as:

\beq
  \label{eq:cal}
  \left(\frac{\sigma_E}{E}\right)^2 = \left(\frac{a}{\sqrt{E}}\right)^2
  + \left(\frac{b}{E}\right)^2 + c^2 \ .
\eeq

\begin{figure}[t]
  \setlength{\unitlength}{1cm}
  \begin{minipage}[t]{6cm}
    \includegraphics[width=6cm]{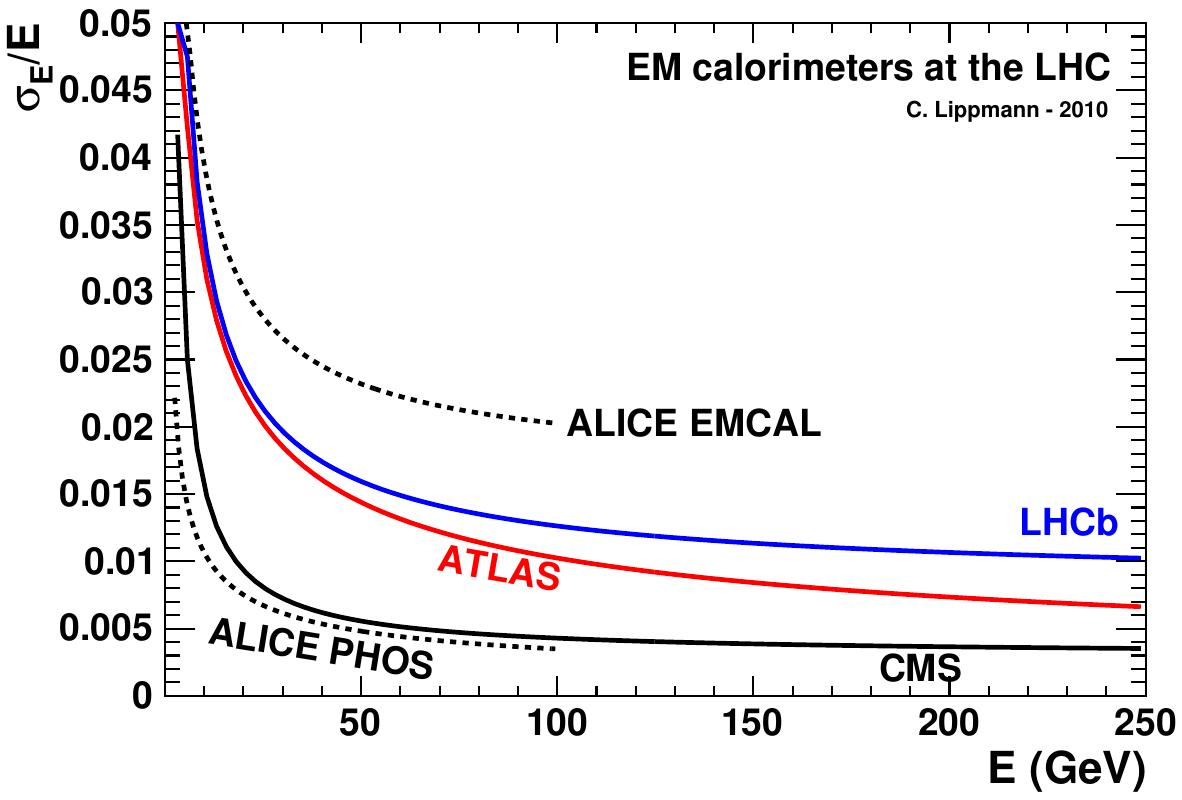}
  \end{minipage}
  \hfill
  \begin{minipage}[t]{6cm}
    \includegraphics[width=6cm]{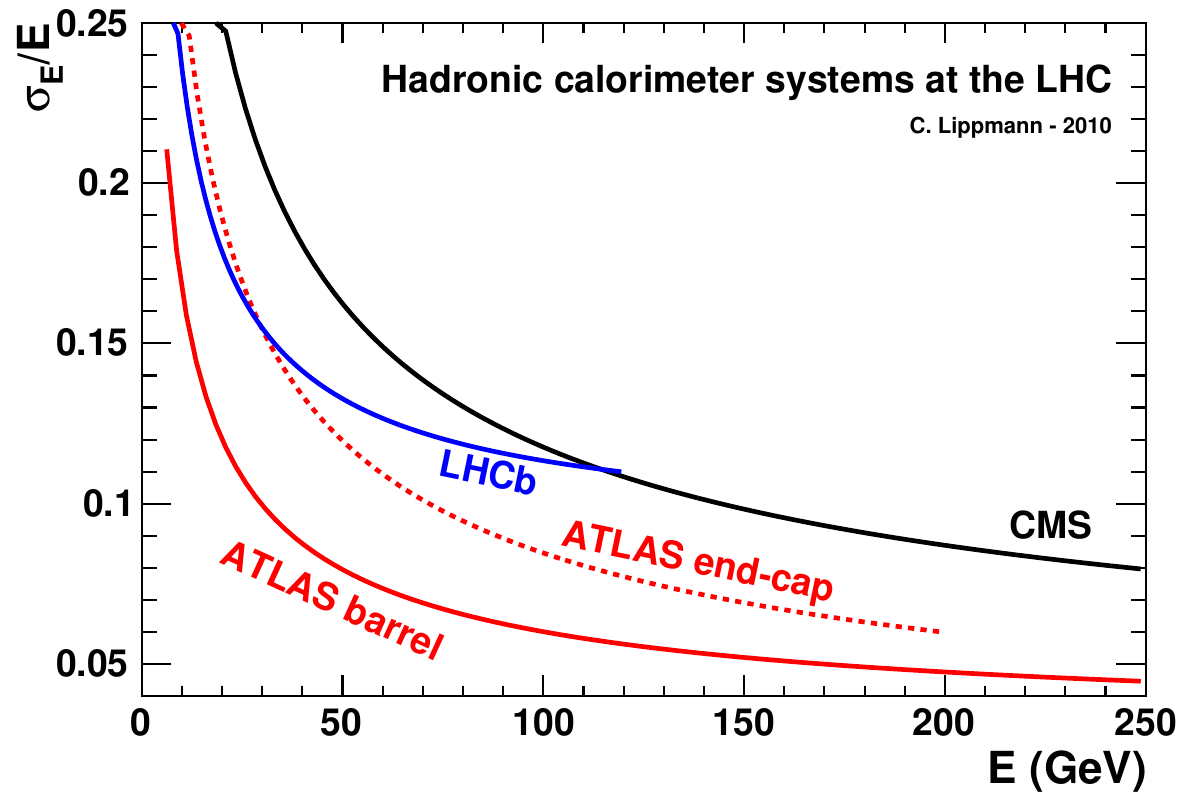}
  \end{minipage}
  \hfill
  \begin{minipage}[b]{12cm}
    \caption{Comparison of the relative energy resolutions (as given by
      Eq.~\ref{eq:cal}) of the different EM calorimeters (left image)
      and hadronic calorimeters (right image) at the LHC
      experiments. The values of the parameters $a$, $b$ and $c$ were
      in all cases determined by fits to the data from beam tests and
      are given in the descriptions of the different experiments in
      Sects. \ref{sec:atlascms}, \ref{sec:alice} and
      \ref{sec:lhcb}. In case of the ATLAS and CMS hadronic
      calorimeters the resolutions of the whole systems combining EM
      and hadronic calorimeters are shown.}
    \label{fig:cal}
  \end{minipage}
\end{figure}

\noindent The first term takes into account the stochastic
fluctuations and limits the low energy performance. The second
term is due to electronics noise. The third, constant term takes into
account detector uniformities and errors in the calibration. This term
limits the calorimeter performance at high energies.
Fig.\,\ref{fig:cal} shows a comparison of the relative resolutions for
the EM and hadronic calorimeters of the different LHC experiments.

Calorimeters may be broadly classified into one of two types: sampling
calorimeters and homogeneous calorimeters. Sampling calorimeters
consist of layers of a dense passive absorber material interleaved with
active detector layers. In homogeneous calorimeters on the other hand
the absorber also acts as the detection medium.

Photons, electrons and positrons deposit all their energy in the EM
calorimeter. Their showers are indistinguishable, but an
electron\footnote{The term ``electron'' can sometimes refer to both
  ``electron'' and ``positron'' in this article. The usage should be
  clear from the context. The same is true for the muon and its 
  anti-particle.} can be identified by the existence of a track in
the tracking system that is associated to the shower. In this case the
energy deposit must match the momentum measured in the tracking
system. Hadrons on the other hand deposit most of their energy
in the hadronic calorimeter (part of it is also deposited in the EM
calorimeter). However, the individual members of the families of
charged and neutral hadrons can not be distinguished in a
calorimeter.

\subsubsection{Muon system}

Muon systems at the LHC are described in detail in Ref.~\cite{muon}.
The muon differs from the electron only by its mass, which is around a
factor 200 larger. As a consequence, the critical energy $E_c$ (the
energy for which in a given material the rates of energy loss through
ionization and bremsstrahlung are equal) is much larger for muons: it
is around 400\,GeV for muons on copper, while for electrons on
copper\footnote{For electrons an approximation for the critical
  energy is given by $E_c = \frac{800}{Z+1.2}$\,MeV, where $Z$ is
  the charge number of the material.} it is only around
20\,MeV. As a consequence, muons do in general not produce
electromagnetic showers and can thus be identified easily by their
presence in the outermost detectors, as all other charged particles
are absorbed in the calorimeter system.

\subsubsection{Other particles}

Neutrinos do in general not interact in a particle detector of the
sort shown in Fig.~\ref{fig:exp}, and therefore escape undetected.
However, their presence can often be inferred by the momentum
imbalance of the visible particles. In electron-positron colliders it
is usually possible to reconstruct the neutrino momentum in all three
dimensions and its energy.

Quark flavor tagging identifies the flavor of the quark that is the
origin of a jet. The most important example is B-tagging, the
identification of beauty quarks. Hadrons containing beauty quarks can
be identified because they have sufficient lifetime to travel a small
distance before decaying. The observation of a secondary vertex a
small distance away from the collision point indicates their
presence. For this the information of a high-precision tracking system
around the collision point is used. Such vertex tracking detectors are
described in detail in Ref.~\cite{vertex}. Also Tau leptons, with a mean
lifetime of 0.29\,ps, fly a small distance (about 0.5\,mm) before
decaying. Again, this is typically seen as a secondary vertex, but
without the observation of a jet.

\begin{figure}[t]
  \begin{center}
    \includegraphics[width=8cm]{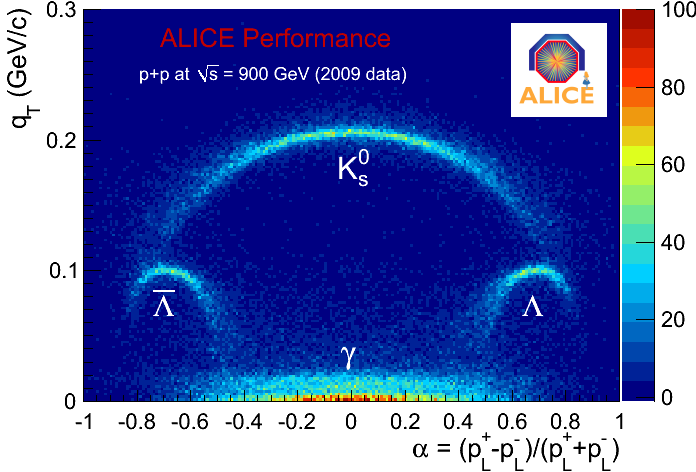}
    \caption{{\it Armenteros-Podolanski plot} from the ALICE
      experiment using data from $\sqrt{s}=900$\,GeV proton
      collisions. The different V$_0$ particles can be separated using the
      kinematics of their decay products. The orientation of the decay
      is described with respect to the momentum vector of the $V_0$.
      $p^\pm_L$ are the longitudinal momenta of the positively and
      negatively charged decay products with respect to the V$_0$
      particle's direction. $q_T$ represents the transverse component
      of the momentum of the positive decay product.}
    \label{fig:amenteros}
  \end{center}
\end{figure}

K$_S^0$, $\Lambda$ and $\bar{\Lambda}$ particles are collectively
known as V$^0$ particles, due to their characteristic decay vertex,
where an unobserved neutral strange particle decays into two observed
charged daughter particles:~e.g. K$_S^0 \rightarrow \pi^+\,\pi^-$ and
$\Lambda \rightarrow $p\ $\pi^-$. V$^0$ particles can be identified
from the kinematics of their positively and negatively charged decay
products (see Fig.\ref{fig:amenteros})~\cite{amenteros}.

\subsection{PID by mass determination}
\label{sec:chPID}

\begin{figure}[t]
  \centering
  \includegraphics[width=12cm]{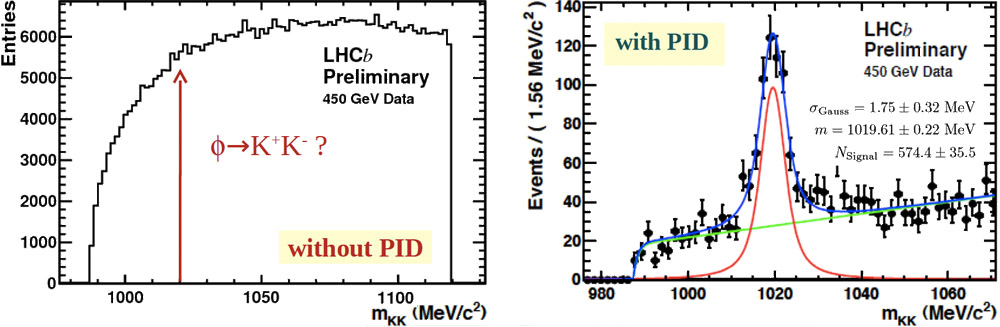}
  \caption{Demonstration of the power of PID by mass determination,
    using the example of the $\phi \rightarrow$\ K$^+$K$^-$ decay
    measured with the LHCb RICH system (preliminary data from
    $\sqrt{s}=900$\,GeV p--p collisions~\cite{lhcbmoriond}). The left
    image shows the invariant mass obtained from all combinations of
    pairs of tracks without PID. The right image shows how the $\phi$
    meson signal appears when tracks can be identified as kaons.}
  \label{fig:richpid}
\end{figure}

The three most important charged hadrons (pions, kaons and protons)
and their antiparticles have identical interactions in an experimental
setup as the one shown in Fig.~\ref{fig:exp} (charge deposit in
the tracking system and hadronic shower in the calorimeter). Moreover,
they are all effectively stable. However, their identification can be
crucial, in particular for the study of hadronic decays. The possible
improvement in the signal-to-background ratio when using PID is
demonstrated in Fig.~\ref{fig:richpid}, using the example of
the $\phi \rightarrow$\ K$^+$K$^-$ decay.

In B physics, the study of hadrons containing the beauty quark,
different decay modes usually exist, and their individual properties
can only be studied with an efficient hadron identification, which
improves the signal-to-background ratio (most tracks are pions from
many sources).

PID is equally important in heavy-ion physics. An example is the
measurement of open charm (and open beauty), which allows to
investigate the mechanisms for the production, propagation and
hadronization of heavy quarks in the hot and dense medium produced in
the collision of heavy ions. The most promising channel is 
D$^0 \rightarrow \mbox{K}^- \pi^+$. It requires a very efficient PID,
due to the small signal-to-background ratio.

In order to identify any stable charged particle, including charged
hadrons, it is necessary to determine its charge $ze$ and its mass
$m$. The charge sign is obtained from the curvature of the particle's
track. Since the mass can not be measured directly, it has to be
deduced from other variables. These are in general the momentum $p$
and the velocity $\beta=v/c$, where one exploits the basic relationship

\beq
  \label{eq:basic}
  p = \gamma m v \quad \rightarrow \quad m = \frac{p}{c\beta\gamma}\ .
\eeq

\noindent Here $c$ is the speed of light in vacuum and
$\gamma=(1-\beta^2)^{-1/2}$ is the relativistic Lorentz factor. The
resolution in the mass determination is

\beq
  \label{eq:merror}
  \left(\frac{\dd m}{m}\right)^2 = \left(\frac{\dd p}{p}\right)^2
    + \left(\gamma^2 \frac{\dd \beta}{\beta}\right)^2 .
\eeq

\noindent Since in most cases $\gamma \gg 1$, the mass resolution is
determined mainly by the accuracy of the velocity measurement,
rather than the momentum determination. 

The momentum is obtained by measuring the curvature of the track in
the magnetic field. The particle velocity is obtained by means of one
of the following methods:

\begin{enumerate}
\item measurement of the energy deposit by ionization,
\item time-of-flight (TOF) measurements,
\item detection of Cherenkov radiation or
\item detection of transition radiation.
\end{enumerate}

\noindent Each of these methods provide PID not only for charged
hadrons, but also for charged leptons. The small obstacle of muons and
pions not being well separated due to $m_\mu \approx m_\pi$ can
luckily be circumnavigated, since muons can be easily identified by
other means.

\begin{figure}[t]
  \centering
  \includegraphics[width=10cm]{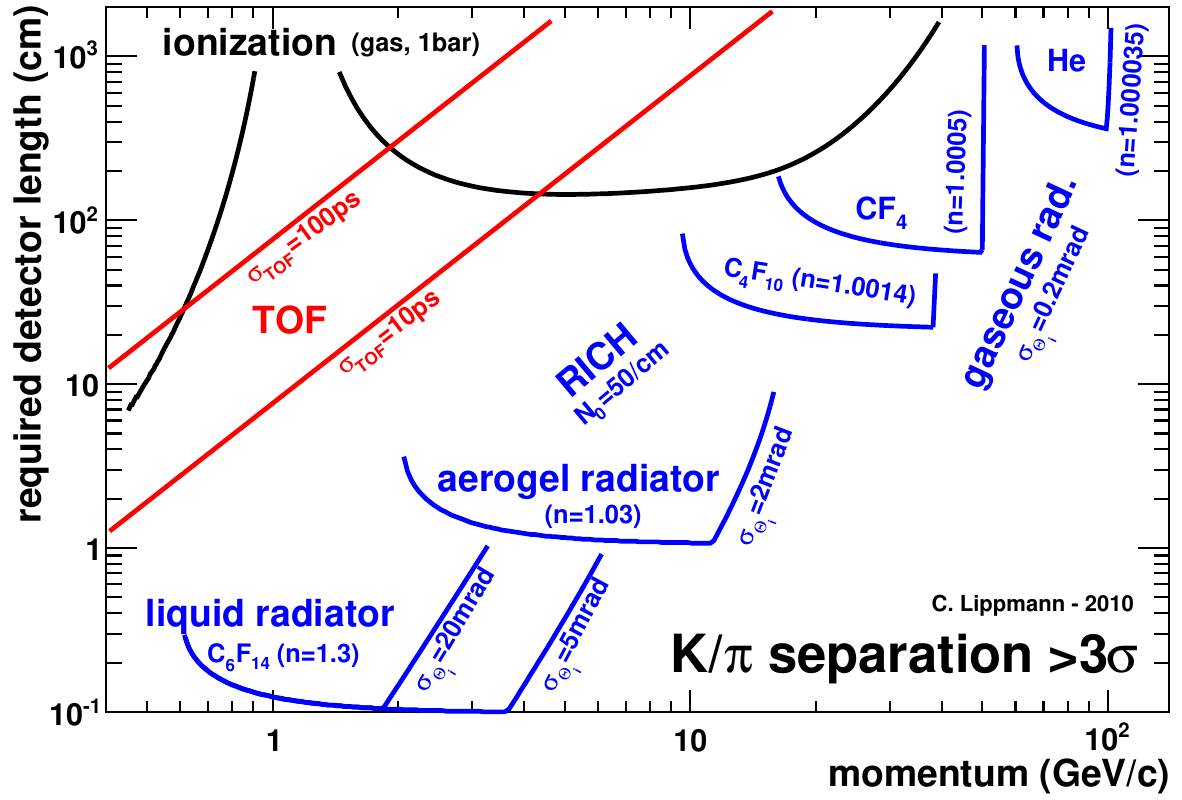}
  \caption{Approximate minimum detector length required to achieve a
    K/$\pi$ separation of $n_\sigma\geq 3\,\sigma$ with three different PID
    techniques. For the energy loss technique we assume a gaseous
    detector. For the TOF technique, the detector length represents the
    particle flight path over which the time-of-flight is
    measured. For the Cherenkov technique only the radiator thickness
    is given. The thicknesses of an expansion gap and of the readout
    chambers have to be added.}
  \label{fig:summary}
\end{figure} 

The use of these methods is restricted to certain momentum ranges.
For a given momentum range, the {\it separation power} can be used to
quantify the usability of a technique. It defines the significance of
the detector response $R$. If $R _A$ and $R_B$ are the mean values of
such a quantity measured for particles of type $A$ and $B$,
respectively, and $\langle\sigma_{A,B}\rangle$ is the average of the
standard deviations of the measured distributions, then the separation
power $n_\sigma$ is given by

\beq
  \label{eq:sep}
  n_\sigma = \frac{R _A - R_ B}{\langle\sigma_{A,B}\rangle} \ .
\eeq

\noindent A summary of the momentum coverage and required detector
lengths using the example of K/$\pi$ separation with the requirement
$n_{\sigma}\geq 3$ is given in Fig.~\ref{fig:summary}.

Naturally, when choosing a PID technique, also other features have to
be considered besides the separation power. In practice, these often
include luminosity and event rates, size and space requirements,
accessibility, multiple scattering in the used materials,
compatibility with other detector subsystems and geometrical
coverage.

\section{Overview of the large LHC experiments}
\label{sec:overview}

The four main LHC experiments are the largest current collider
experiments and integrate todays state-of-the-art detector
technology, in particular with respect to PID. This section gives an
overview on the four experiments, with emphasis on the way the
underlying physics program influences the experiment design with
respect to PID.

\subsection{ATLAS and CMS}
\label{sec:atlascms}

ATLAS~\cite{atlas} (A Toroidal Large AparatuS) and CMS~\cite{cms}
(Compact Muon Solenoid) are often called ``general purpose'' particle
physics experiments, since they aim at uncovering any new phenomena
appearing in proton--proton (p--p) collisions at the new energy domain
that is now probed at the LHC. The main focus of the experiments is the
investigation of the nature of the electroweak symmetry breaking and
the search for the Higgs boson. The experimental study of the Higgs 
mechanism is expected to shed light on the mathematical consistency of
the Standard Model (SM) of particle physics at energy scales above
$\sim$1\,TeV. As a matter of fact, various alternatives to the SM
foresee new symmetries, new forces, and new constituents. Furthermore,
there are high hopes for discoveries that could lead towards a unified
theory (supersymmetry, extra dimensions, \dots).

\subsubsection{Requirements}

The discovery and study of the Higgs boson is the benchmark process
that influenced the design of the two experiments. Electrons, muons and
photons are important components of its possible physics signatures.
Assuming a low mass Higgs boson ($m_H<150$\,GeV/c$^2$), the
predominant decay mode into hadrons is difficult to detect due to QCD
backgrounds. In that case an important decay channel is
$H \rightarrow \gamma \gamma$. For masses above 130\,GeV, the most
promising channel to study the properties of the Higgs boson is
$H \rightarrow ZZ^{(*)} \rightarrow  4\,l$. It is called
``gold-plated'' in the particular case in which all four leptons
are muons, due to the relative ease in detecting muons. The ATLAS and
CMS detectors have been optimised to cover the whole spectrum of
possible Higgs particle signatures. In summary, these are the
requirements:

\begin{enumerate}
\item large acceptance in pseudo-rapidity\footnote{The pseudo-rapidity
    is defined as $\eta = -\ln [tan(\Theta/2)]$, where $\Theta$
    is the polar angle between the charged particle direction and the
    beam axis.} and almost full
  coverage in azimuthal angle (the angle around the beam direction);
\item good identification capabilities for isolated high transverse
  momentum\footnote{The transverse component of the momentum is the
    one in the plane that is perpendicular to the beam direction.}
  photons and electrons;
\item good muon ID and momentum resolution over a wide range of
  momenta and angles. At highest momenta (1\,TeV) a transverse
  momentum resolution $\sigma_{p_T}/p_T$ of the order 10\% is
  required.
\end{enumerate}

To maximize the integrated luminosity for the rare processes
that are the main interest of the experiments, proton bunches
will be brought to collision every 25\,ns. Strong focusing of the
beams helps to increase the luminosity to
$10^{34}$\,cm$^{-2}$s$^{-1}$. The consequences are a large number
of overlapping events per proton bunch crossing and extreme particle
rates. For the experiments these pose major challenges.

\subsubsection{Setup}

\begin{figure}[p]
  \centering
  \includegraphics[width=12cm]{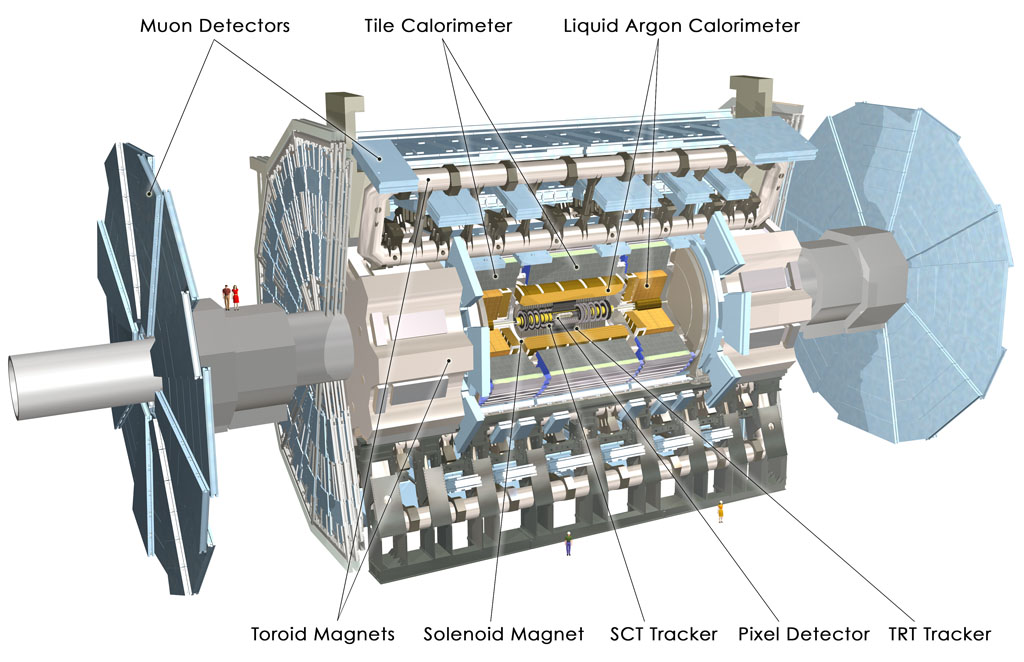}
 \caption{Perspective view of the ATLAS detector~\cite{atlas}. The
   dimensions are 25\,m in height and 44\,m in length, the overall
   weight of the detector is approximately 7000 tonnes.}
  \label{fig:atlas}
\end{figure}
 
\begin{figure}[t]
  \centering
  \includegraphics[width=12cm]{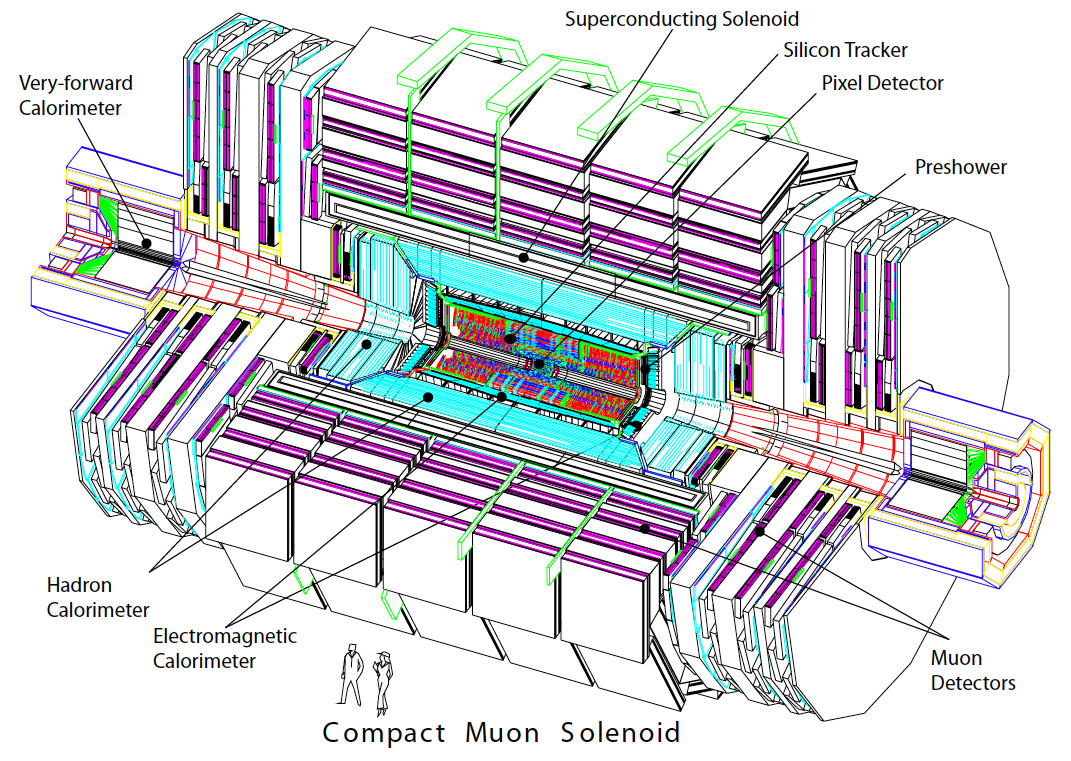}
  \caption{Perspective view of the CMS detector~\cite{cms}. The
    dimensions are 14.6\,m in height and 21.6\,m in length. The
    overall weight is approximately 12\,500 tonnes.}
  \label{fig:cms}
\end{figure}

The setup of ATLAS and CMS is in general quite similar, following the
``traditional'' setup as described in Sect. \ref{sec:classic}. However,
in the implementation of some of the components quite different
choices were made. The main similarities and some differences can be
seen in Tab. \ref{tab:atlascms}. Schematic images of the two detectors
are shown in Figs. \ref{fig:atlas} and \ref{fig:cms}.

\begin{table}[p]\footnotesize
  \caption{Overview on the technologies chosen for the ATLAS and CMS
    tracking systems, EM and hadronic calorimeters and muon systems.
    Their acceptances in pseudo-rapidity $\eta$ are given as
    well. Abbreviations are explained in the text.}
  \label{tab:atlascms}
  \begin{center}
    \begin{tabular}{|l||c|c|c|c|}
      \hline
      {\bf  Detector} & \multicolumn{2}{|c|}{\bf technology} &
      \multicolumn{2}{|c|}{\bf $\eta$-coverage} \\
      {\bf  component} & {\bf ATLAS} & {\bf CMS} & {\bf ATLAS} &
      {\bf CMS} \\ 
      \hline
      \hline
      {\bf Tracking} & \multicolumn{2}{|c|}{Si pixel detector (3
        layers)} & \multicolumn{2}{|c|}{} \\
      \cline{2-3}
     & \multicolumn{2}{|c|}{Si strip detector} &
      \multicolumn{2}{|c|}{$|\eta |<2.5$} \\
      & SCT (4 layers) & (10 layers)  & \multicolumn{2}{|c|}{} \\
      \cline{2-3}
      & TRT (straw tubes) & - & \multicolumn{2}{|c|}{} \\
      \hline 
     {\bf EM Cal} & Sampling & Homogeneous & $|\eta|<3.2$ & $|\eta|<3.0$ \\
      & (Pb / LAr) & (PbWO$_4$ crystals) & & \\
      \hline 
      {\bf H Cal} & Sampling & Sampling & & \\
      & (Barrel: Iron / Scint. & (Brass / Scint.) & $|\eta|<3.2$ &
      $|\eta|<3.0$ \\
      & End-caps: Cu / LAr) & & & \\
     \hline 
     {\bf Muon} & CSC (inner plane) & CSC & $2<|\eta|<2.7$ &
      $0.9<|\eta|<2.4$\\
     \cline{2-5}
      {\bf (Tracking)} & MDT & DT & $|\eta|<2.7$\,($2.0$ & $|\eta|<1.2$ \\
      & & & for inner plane) & \\
      \hline
     {\bf Muon} & \multicolumn{2}{|c|}{Bakelite RPC} &
      $|\eta|<1.05$ & $|\eta|<1.6$\\
      \cline{2-5}
     {\bf (Trigger)} & TGC & - & $1.05<|\eta|<2.7$ & - \\
      \hline
  \end{tabular}
  \end{center}
\end{table}
 
\subsubsection{Tracking and muon systems}

As the innermost component, the tracking systems of ATLAS and CMS
are embedded in solenoidal magnetic fields of 2\,T and 4\,T,
respectively. In both cases the tracking systems consist of silicon
pixel and strip detectors. ATLAS includes also a Transition
Radiation Tracker (TRT) based on straw tubes, which provides
also electron ID~\cite{trds}.

The global detector dimensions are defined by the large muon
spectrometers, which are designed to measure muon momenta with
extremely high accuracy~\cite{muon}. While the muon system of ATLAS is
designed to work independently of the inner detector, in CMS the
information from the tracking system is in general combined with that
from the muon system.

The ATLAS muon system uses eight instrumented air-core toroid coils,
providing a magnetic field mostly orthogonal to the muon trajectories,
while minimising the degradation of resolution due to multiple
scattering~\cite{atlas}. For high-precision tracking, the magnets are
instrumented with Monitored Drift Tubes (MDT) and, at large
pseudo-rapidities, Cathode Strip Chambers (CSC). By measuring
muon tracks with a resolution $\leq$50\,$\mu$m, a standalone
transverse momentum resolution $\sigma(p_T)/p_T\approx3$\% is achieved
for $p_T=100$\,GeV/c, while at $p_T=1$\,TeV/c it is around 10\%. Only
below 200\,GeV/c the combination of the information from the muon and
tracking systems helps, keeping the resolution below 4\%. As a
separate trigger system and for second coordinate measurement,
Resistive Plate Chambers (RPCs) are installed in the barrel, while
Thin Gap Chambers (TGCs) are placed in the end-caps, where particle
rates are higher.

The CMS design relies on the high bending power (12\,Tm) and momentum
resolution of the tracking system, and uses an iron yoke to increase
its magnetic field~\cite{cms}. With the field parallel to the LHC beam
axis, the muon tracks are bent in the transverse plane. The iron yoke is
instrumented  with aluminum Drift Tubes (DT) in the barrel and CSCs in
the end-cap region. Due to the iron yoke, the momentum resolution of
the CMS muon system is dominated by the multiple scattering. As a
consequence, the requirements on spatial resolution are somewhat looser
($\sim$70\,$\mu$m) than in ATLAS. The standalone muon momentum
resolution is $\sigma(p_T)/p_T=9$\% for $p_T\leq 200$\,GeV/c and
15 to 40\% at $p_T=1$\,TeV/c, depending on $|\eta|$. Including the
tracking system improves the result by an order of magnitude for low
momenta. At 1\,TeV the contribution of both measurements together
yield a momentum resolution of about 5\%. The DT and CSC subsystems
can each trigger on muons with large transverse momentum. However, for
the full LHC luminosity, faster trigger chambers are needed to
associate the detected muons to the right crossing of proton bunches
in the LHC. RPCs are used throughout the whole CMS muon system for
that purpose.

\subsubsection{Particle identification}

Electrons, hadrons and neutral particles are identified in the
calorimeter systems, while muons are identified in the large muon
systems. The designs of the EM calorimeters and muon systems have
been guided by the benchmark processes of Higgs boson decays into
two photons or into leptons.

The ATLAS EM calorimeter is of sampling type and consists of liquid
Argon (LAr) as detection medium and lead (Pb) absorber plates operated
in a cryostat at 87\,K. The calorimeter depth varies from 22 to
38\,$X_0$, depending on the pseudo-rapidity range. In beam tests~\cite{atlas}  
the relative energy resolution after subtraction of the noise term was
found to be $(\sigma_E/E)^2=(0.1/\sqrt{E\,(\mbox{GeV})})^2+0.007^2$.
The CMS EM calorimeter on the other hand is a homogeneous calorimeter
made from lead tungstate (PbWO$_4$) crystals with a length that
corresponds to around $25 \,X_0$. The relative energy resolution was
measured in beam tests\footnote{The measurement was done for
  electrons, requiring their impact point to lie in the center of the
  $3\times 3$ crystals used to evaluate the energy deposit. Without
  this  requirement the relative resolution is slightly worse but
  still meeting the design goal of better than 0.5\% for
  $E>100$\,GeV~\cite{cms}.} as $(\sigma_E/E)^2 =
(0.028/\sqrt{E\,(\mbox{GeV})})^2 + (0.125/E($\,GeV$))^2 + 0.003^2$.

The ATLAS hadronic calorimeter is made from a Barrel and two end-cap
modules. The barrel part is made from steel and plastic scintillator
tiles. The end-caps are made from copper plates and use LAr as active
medium. In beam tests of the combined EM and hadronic calorimeter
systems the relative energy resolution~\cite{atlas} for pions was
found to be $(\sigma_E/E)^2 = (0.52/\sqrt{E\,(\mbox{GeV})})^2 +
(0.016/E($\,GeV$))^2 + 0.03^2$ in the barrel\footnote{For the barrel
  hadron calorimeter standalone the relative energy resolution for
  pions is $(\sigma_E/E)^2 = (0.564/\sqrt{E\,(\mbox{GeV})})^2 +
  0.055^2$.} and $(\sigma_E/E)^2 \approx
(0.84/\sqrt{E\,(\mbox{GeV})})^2$ in the end-caps. The CMS 
hadronic calorimeter is made from brass and plastic scintillator
tiles. The resulting energy resolution\,\cite{cmshcal} for the
combined system of EM and hadronic calorimeters and for pions
is $(\sigma_E/E)^2 = (1.12/\sqrt{E\,(\mbox{GeV})})^2 +
0.036^2$.

The silicon detectors of the ATLAS and CMS tracking systems offer the
possibility of hadron ID at low momenta (a few hundred MeV/c) via
ionization measurements. For electron ID at momenta up to 25\,GeV/c
ATLAS also takes into account the information from the TRT, namely
large energy deposits by electrons due to transition radiation (for
details see Ref.\,\cite{trds}).

\subsection{ALICE}
\label{sec:alice}

ALICE~\cite{alice} (A Large Ion Collider Experiment) is the dedicated
heavy-ion experiment at the LHC. The LHC can collide lead (Pb)
nuclei at center-of mass energies of $\sqrt{s_{NN}}=2.76$ and
5.5\,TeV. This leap to up to 28 times beyond what is presently
accessible will open up a new regime in the experimental study of
nuclear matter. The aim of ALICE is to study the physics of strongly
interacting matter at the resulting extreme energy densities and to
study the possible formation of a quark-gluon plasma. Many observables
meant to shed light on the evolution of the quark-gluon plasma depend
on PID. The most natural example is particle spectra, from which the
freeze-out parameters (such as the kinetic and chemical freeze-out
temperature and the collective flow velocity) can be extracted. Besides 
pions, kaons and protons, also resonances such as the $\phi$ meson can
be used for such analysis. Actually, any modification in the mass and
width of the $\phi$ meson could be a signature for partial chiral
symmetry restoration, and thus, creation of a quark-gluon plasma.
The $\phi$ resonance is typically identified via its hadronic decay
channel $\phi \rightarrow K^+ K^-$, which makes a good hadron
ID important. Leptonic channels are very important as well and thus
extremely good lepton ID is needed.

ALICE also studies p--p collisions at $\sqrt{s}=7$ and 14\,TeV. These
data are important as a reference for the heavy-ion data. Moreover,
due to the very low momentum cut-off ($\sim$0.1\,GeV/c), ALICE can
contribute in physics areas where it complements the other LHC
experiments. To name an example, the measurement of charm and beauty
cross sections is possible down to very low momenta. Moreover,
proton physics at high multiplicities is easily accessible via the
multiplicity trigger from the Silicon Pixel Detector (SPD).

\subsubsection{Requirements}

As compared to p--p collisions, the charged track multiplicities
in Pb--Pb collisions are extraordinary. At the time of the
ALICE Technical Proposal, charged particle densities of up to
d$N_\text{ch}$/d$\eta=8000$ were considered for the LHC center-of-mass
energy~\cite{tpalice,ppr1,ppr2} . Including secondaries, this would
amount to 20\,000 tracks in one interaction in the relevant acceptance
region. However, first measurements~\cite{alicePb1} at
$\sqrt{s_{NN}}=2.76$\,TeV yield d$N_\text{ch}$/d$\eta\approx
1600$. When compared to p--p collisions at the LHC, the produced
particles have relatively low momenta:~i.e. 99\% have a momentum
below about 1\,GeV/c. Hadrons, electrons, muons and photons may be
used as probes in order to explore the strongly interacting matter
that is produced in the Pb--Pb collisions. Those probes are inspected
by dedicated PID detectors. In some cases these detectors may have
very limited geometrical acceptance; for some actually not even a coverage
of 2\,$\pi$ in azimuthal angle is necessary. This is an important
difference when compared to general purpose experiments. Nevertheless,
large geometrical acceptance remains important for the main tracking
devices. In summary, these are the requirements for the
ALICE experiment:

\begin{enumerate}
\item reliable operation in an environment of very large charged
  track multiplicities;
\item precision tracking capabilities at very low momenta
  (100\,MeV/c), but also up to 100\,GeV/c;
\item low material budget;
\item low magnetic field ($0.2\leq B\leq 0.5$\,T) in order to be able
  to track low momentum particles;
\item good hadron ID for momenta up to a few GeV/c and electron
  ID up to 10\,GeV/c in the central barrel;
\item good muon ID (in the forward region).
\end{enumerate}

At LHC energies the total cross-section for Pb--Pb collisions will be
of the order of 8\,b. At the maximum expected luminosity of
$10^{27}$\,cm$^{-2}$s$^{-1}$ this corresponds to an interaction rate
of 8\,kHz. Only a fraction of these events will be central collisions
(collisions with small impact parameter) and the aim is to be able to
trigger on at least 200 of such central events per second.

For p--p collisions the goal is to be able to read out the experiment
at rates of at least 1.4\,kHz. Since the specific features of the used
detectors (TPC drift time of almost 100\,$\mu$s, see
Sect.~\ref{sec:alicetpc}) makes it a rather slow detector (at least
compared to the three other large LHC experiments), the
luminosity for p--p  collisions has to be limited~\cite{ppr1}: at
$\sim$$5\times 10^{30}$\,cm$^{-2}$s$^{-1}$, corresponding to an
interaction rate of $\sim$200\,kHz, the integrated luminosity can
be maximised for rare processes. With lower luminosity
($10^{29}$\,cm$^{-2}$s$^{-1}$) statistics for large cross section
observables can be collected and global event properties can be
studied at optimum detector performance.

\subsubsection{Setup}

\begin{figure}[t]
  \centering
  \includegraphics[width=12cm]{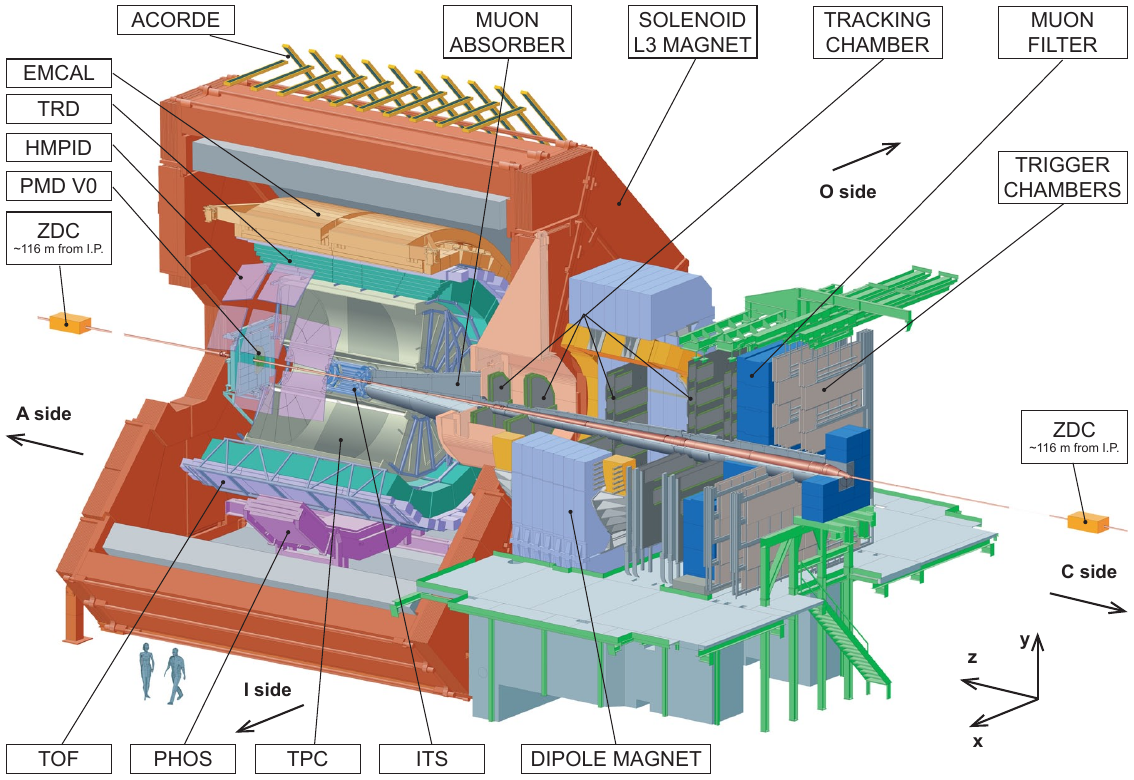}
  \caption{Perspective view of the ALICE detector~\cite{alice}. The
    dimensions are 16\,m in height and 26\,m in length. The overall
    weight is approximately 10\,000 tonnes.}
  \label{fig:alice}
\end{figure}
 
The ALICE experimental setup is shown in Fig. \ref{fig:alice}. ALICE
consists of a central barrel part for the measurement of hadrons,
leptons and photons, and a forward muon spectrometer. The central part
is embedded in a large solenoid magnet that is reused from the L3
experiment at the LEP collider at CERN.
 
A Time Projection Chamber (TPC) was chosen as the main tracking device
in the central barrel, as this is a very high granularity detector
capable of satisfying the requirements given in the previous
section. The TPC surrounds the Inner Tracking System (ITS) which is
optimised  for the determination of primary and secondary vertices
and for precision tracking of low-momentum particles. As the innermost
layer, the Silicon Pixel Detector (SPD) has a key role in the
determination of the vertex position. A unique feature of the SPD is
its capability to generate a prompt trigger based on a programmable,
fast online algorithm. As second and third layers, a Silicon Drift
Detector (SDD) and a Silicon Strip Detector (SSD) complete the ITS,
with two detection planes each. Finally, on the outside of the TPC,
the Transition Radiation Detector (TRD) contributes to the tracking
capability and provides electron ID~\cite{trds}.

\subsubsection{Particle identification}

The strategy for PID in the barrel of ALICE is described in detail in
Ref.~\cite{ppr2}. In a first step, all charged particles from the 
collision are tracked in the ITS, TPC and TRD using parallel Kalman
filtering. The PID procedure is then applied to all reconstructed
tracks that have been successfully associated to a signal in one of
the PID detectors.
 
The TPC, SSD and SDD each provide PID via ionization measurements.
The TPC is described in more detail in Sect.~\ref{sec:alicetpc}. The
TRD is designed for electron ID and is described in
Ref.~\cite{trds}. The time-of-flight (TOF) array provides charged
hadron ID and is described in Sect.~\ref{sec:alicetof}.

Surrounding the TOF detector, there are three single-arm detectors
inside the ALICE solenoid magnet: the Photon Spectrometer (PHOS), the
Electro-Magnetic CALorimeter (EMCAL) and an array of RICH counters
optimised for High-Momentum Particle IDentification (HMPID, see
Sect.~\ref{sec:alicehmpid}). PHOS is a homogeneous EM calorimeter
based on lead tungstate crystals, similar to the ones used by CMS,
read out using Avalanche Photodiodes (APD). It covers 100\,deg in
azimuthal angle and $|\eta|\leq 0.12$ in pseudo-rapidity. Its
thickness corresponds to 20\,$X_0$. The crystals are kept at a
temperature of 248\,K to minimize the contribution of noise to the
energy resolution, optimizing the response for low energies. In a beam
test the relative energy resolution was found to be
$(\sigma_E/E)^2=(0.033/\sqrt{E\,(\mbox{GeV})})^2 +
(0.018/E\,($GeV$))^2 + 0.011^2$.
The EMCAL was proposed as a late addition to the ALICE setup, with the
primary goal to improve the capabilities of ALICE for jet measurements.
The main design criterion was to provide as much EM calorimetry
coverage as possible within the constraints of the existing ALICE
detector systems. The EMCAL is a sampling calorimeter made from
lead absorber plates and scintillators and is positioned
opposite in azimuth to the PHOS, covering 107\,deg in azimuthal angle
and $|\eta|\leq 0.7$ in pseudo-rapidity. The relative energy resolution
of the EMCAL was measured in a test beam and can be parameterised
as $(\sigma_E/E)^2 = (0.113/\sqrt{E\,(\mbox{GeV})})^2 + 0.0168^2$.

The goal of the ALICE muon spectrometer is to study vector mesons
containing heavy quarks, such as $J/\Psi$, $\Psi'$ and members of
the $\Upsilon$ family via the muonic channel. The muon spectrometer 
covers only the limited pseudo-rapidity interval $2.5\leq\eta\leq 4$.
Closest to the interaction region there is a front absorber to
remove hadrons and photons emerging from the collision. Five pairs of
high-granularity detector planes form the tracking system within the
field of a large dipole magnet. Beyond the magnet is a
muon filter (a 120\,cm thick iron wall), followed by two pairs of
trigger chamber planes (RPCs).

\subsection{LHCb}
\label{sec:lhcb}

LHCb~\cite{lhcb} is dedicated to heavy flavor physics. One particular
aim is to look at evidence of new physics in CP violation and rare
decays of beauty and charm hadrons. The level of CP violation in the
SM can not explain the absence of antimatter in our universe. A new
source of CP violation is needed to understand this matter-antimatter
asymmetry, implying new physics. Particles associated with new physics
could manifest themselves indirectly in beauty or charm meson decays
and produce contributions that change the expectations of CP violation
phases. They may also generate decay modes forbidden in the SM.

\subsubsection{Requirements}

A large $b\bar{b}$  production cross section of the order
$\sim$500\,$\mu$b is expected for p--p collisions at
$\sqrt{s}=14$\,TeV. At high energies, the $b\bar{b}$ pairs are
predominantly produced in a forward or backward cone. Separating pions
from kaons in selected B hadron decays is fundamental to the physics
goals of LHCb. A good example is the channel
$B_S^0 \rightarrow D_S^\mp K^\pm$, which has to be separated from the
background from $B_S^0 \rightarrow D_S^- \pi^+$, which is about 15
times more abundant. Moreover, other final states containing electrons,
muons and neutral particles (photons, neutral pions and $\eta$'s) have
to be distinguishable. The requirements for the LHCb detector can be
summarised in this way:

\begin{enumerate}
\item geometrical acceptance in one forward region ($1.9 < \eta < 4.9$); 
\item good hadron ID;
\item good momentum and vertex resolutions and
\item an efficient and flexible trigger system.
\end{enumerate}

In order to maximize the probability of a single interaction per beam
crossing, the luminosity in the LHCb interaction region may be
limited to 2 to $5\times 10^{32}$cm$^{-2}$s$^{-1}$. In
these conditions, one year of LHCb running ($\sim$10$^7$\,s)
corresponds to 2\,fb$^{-1}$ of integrated luminosity and about
$10^{12} $ $b\bar{b}$ pairs produced in the region covered by the
spectrometer.

\subsubsection{Setup}

\begin{figure}[t]
  \centering
  \includegraphics[width=12cm]{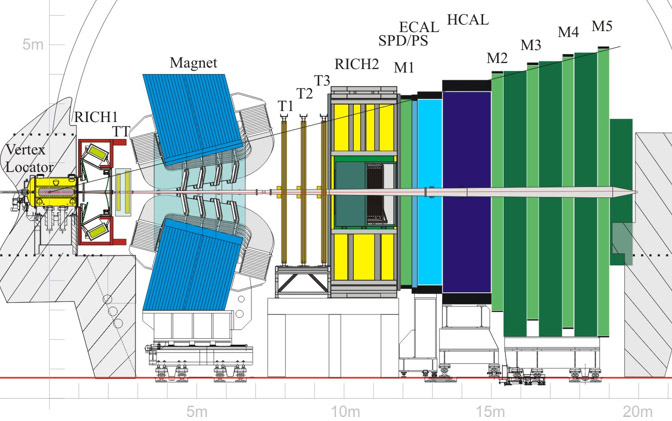}
  \caption{Schematic view of the LHCb detector~\cite{lhcb}.}
  \label{fig:lhcb}
\end{figure}
 
Unlike ATLAS and CMS, LHCb does not have a cylindrical geometry.
Rather, it is laid out horizontally along the beam line, as shown in
Fig. \ref{fig:lhcb}. The tracking system consists of the VErtex
LOcator (VELO) and four planar tracking stations: the Tracker
Turicensis (TT) upstream of the 4\,Tm dipole magnet, and stations
T1-T3 downstream of the magnet. VELO and TT use silicon strip
detectors. In T1-T3, silicon strips are used in the region close
to the beam pipe, whereas strawtubes are employed in the outer
regions. The VELO makes possible a reconstruction of primary vertices
with 10\,$\mu$m (60\,$\mu$m) precision in the transverse (longitudinal)
direction. In this way the displaced secondary vertices, which are a
distinctive feature of beauty and charm hadron decays, may be
identified. The overall performance of the tracking system enables the
reconstruction of the invariant mass of beauty mesons with resolution
$\sigma_m \approx 15$ to $20$\,MeV/c$^2$, depending on the channel.

\subsubsection{Particle identification}

LHCb in general looks like a slice out of a ``traditional'' experiment
as described in Sect.~\ref{sec:classic}, apart from the two RICH
detectors providing hadron ID. The RICH detectors are described
in more detail in Sect. \ref{sec:lhcbrich}. An EM calorimeter and a
hadron calorimeter provide the identification of electrons, hadrons
and neutral particles (photons and $\pi_0$) as well as the measurement
of their energies and positions. The EM calorimeter is a rectangular
wall constructed out of lead plates and scintillator tiles. The total
thickness corresponds to 25\,$X_0$. In a beam test it was found that
the relative energy resolution follows $(\sigma_E/E)^2 =
(0.094/\sqrt{E\,(\mbox{GeV})})^2 + (0.145/E\,($GeV$))^2 + 0.0083^2$.
The hadronic calorimeter consists of iron and scintillator tiles with
a relative energy resolution of
$(\sigma_E/E)^2 = (0.69/\sqrt{E\,(\mbox{GeV})})^2 + 0.09^2$,
measured with a prototype in a beam test. Finally, the muon system is
designed to provide a fast trigger on high momentum muons as well as
offline muon indentification for the reconstruction of muonic final
states and beauty flavor tagging. It consists of five stations (M1-M5)
equipped mainly with Multi Wire Proportional Chambers (MWPCs). For the
innermost region of station M1, which has the highest occupancy,
Gas Electron Multipliers (GEMs)~\cite{gem} are used. For a muon PID
efficiency of 90\% the misidentification rate is only about 1.5\%.

\section{Ionization measurements}
\label{sec:dedx}

Ionization of matter by charged particles is the primary mechanism
underlying most detector technologies. The characteristics of this
process, along with the momentum measurement, can be used to
identify particles.

\subsection{Energy loss and ionization}
\label{sec:dedxdedx}

When a fast charged particle passes through matter, it undergoes a
series of inelastic Coulomb collisions with the atomic electrons of
the material. As a result, the atoms end up in excited or ionised
states, while the particle loses small fractions of its kinetic
energy. The average energy loss per unit path length
$\langle\dd E/\dd x\rangle$ is transformed into the average number
of electron/ion pairs (or electron/hole pairs for semiconductors)
$\langle N_I\rangle$ that are produced along the length $x$ along the
particle's trajectory~\cite{blumrolandi}:

\beq
  \label{eq:w}
  x \left\langle\frac{\dd E}{\dd x}\right\rangle=\langle N_I\rangle\,W\ ,
\eeq

\noindent where $W$ is the average energy spent for the creation of
one electron/ion (electron/hole) pair. $W$ exceeds the ionization
energy $E_I$ (or the band gap energy $E_g$ for a solid) of the
material, because some fraction of the energy loss is dissipated by
excitation, which does not produce free charge carriers. Typical
values of $W$ lie around 30\,eV for gases, being constant for incident
particles with relativistic velocities ($\beta\approx 1$), but
increasing for low velocities. For semiconductors the $W$ values are
roughly proportional to the band gap energy:

\beq
  \label{eq:bandgap}
  W = 2.8\,E_g+0.6\,\mbox{eV}
\eeq

\noindent and are much lower than for gases~\cite{spieler}:~e.g. on
average 3.6\,eV in silicon and 2.85\,eV in germanium. Consequently,
the ionization yield in semiconductor detectors is much larger than in
gaseous devices.

The interactions of the charged particle with the atomic electrons
can be modeled in terms of two components: primary and secondary
interactions. In primary interactions direct processes between the
charged particle and atomic electrons lead to excitation or
ionization of atoms, while secondary processes involve subsequent
interactions. The primary interactions can be characterised by the
Rutherford cross-section (with the energy dependence
$\dd\sigma/\dd E\propto E^{-2}$) for energies above the highest
atomic binding energy, where the atomic structure can be ignored. In
this case the particle undergoes elastic scattering on the atomic
electrons as if they were free. According to the steeply falling
Rutherford spectrum most of the primary electrons emitted in such
collisions have low energy. However, a significant probability for
producing primary electrons with energies up to the kinematic limit
for the energy transfer $E_{max}$ exists. $E_{max}$ is given by

\beq
  \label{eq:emax}
  E_{max} = \frac{2\beta^2\gamma^2m_e c^2}{1+x^2+2\gamma x} \ ,
\eeq

\noindent where $m_e$ is the electron mass, $x=m_e/m$ and $m$
is the mass of the incident particle. In such collisions, characterised
by a very small impact parameter, the energy transfered to the
electron will be larger than $E_I$ (or $E_g$) and the resulting
{\it $\delta$-rays} or {\it knock-on electrons} produce additional
ionization in secondary interactions. {\it $\delta$-rays} can even
leave the sensitive volume of the detector, but a magnetic field may
force them to curl up close to the primary charged particle's
track. In this case they will contribute to a measurement of the
deposited energy.

In collisions with large impact parameter the atomic electrons receive
much less energy, which is used for excitation without the creation
of free charges. However, in gases tertiary ionization by collisions
of an atom in an excited state with other atoms may be important
(Penning ionization)~\cite{penning}.

\subsection{Velocity dependence}
\label{sec:dedxbeta}

The first calculation for the average energy loss per unit track length
based on the quantum mechanical principles of the scattering theory
was introduced by Hans Bethe in 1930 and 1932~\cite{bethe1,bethe2}.
The well known {\it Bethe-Bloch} formula is modified to yield the
{\it restricted}  (average) energy loss by neglecting higher energy
$\delta$-electrons through the introduction of an upper limit for the
energy transfer in a single collision $E_{cut}$~\cite{blumrolandi}:

\beq
  \label{eq:bb}
  \left\langle\frac{\dd E}{\dd x}\right\rangle \propto
  \frac{z^2}{\beta^2}
  \left(\log\frac{\sqrt{2m_e c^2 E_{cut}}\,\beta\gamma}{I} -
    \frac{\beta^2}{2} - \frac{\delta}{2} \right) \ .
\eeq

\noindent Here $ze$ is the charge of the incident particle and $I$ is the
effective excitation energy of the absorber material\footnote{For
  elements the excitation energy (in eV) can be approximately calculated
  as $I\sim 13.5\,Z$ for $Z\leq 14$ and $I\sim 10\,Z$ for $Z>14$,
  where $Z$ is the atomic number of the absorber material. A graphical
  presentation of measured values is given in Ref.~\cite{pdg}.}
measured in eV. $\delta$ is the density effect correction to the
ionization energy loss, which was calculated for the first time by
E.~Fermi in 1939~\cite{fermi}. It is much larger for liquids or solids
than for gases and is usually computed using a parameterization by
R.~M.~Sternheimer~\cite{pdg}. Typical values for $E_{cut}$ depend on
the strength of the magnetic field and lie in the range 10 to
100\,keV.

Eq.~\ref{eq:bb} is valid for electrons and also for heavier particles. 
In the low energy region (i.e. $\beta\gamma<0.5$) the average
energy loss decreases like $1/\beta^2$ and reaches a broad
{\it minimum of ionisation} ({\it MIP region}) at $\beta\gamma \approx
4$. For larger values of $\beta\gamma$ the average energy loss begins
to rise roughly proportional to $\log(\beta\gamma)$
({\it relativistic rise}) with a strength defined by the mean
excitation energy $I$. The rise is reduced at higher momenta by the
density effect correction $\delta$. A remaining relativistic rise
would be due to (rare) large energy transfers to a few
electrons. Since these events are removed in Eq.~\ref{eq:bb} by the
introduction of $E_{cut}$, the restricted average energy loss
approaches a constant value (the {\it Fermi  plateau}) as
$\beta\gamma$ increases. For solids, due to a stronger density effect,
the Fermi plateau is only a few percent above the  minimum.

\subsection{Straggling functions}
\label{sec:straggling}

For a given particle, the actual value of the energy loss $\Delta$
over a given path length $x$ is governed by statistical fluctuations
which occur in the number of collisions (a Poissonian distribution)
and in the energy transferred in each collision (a distribution
$\sim$1/E$^2$). The resulting distribution of the energy loss $F(x,
\Delta)$ is called {\it straggling function}. One particular way to
calculate this distribution using specific assumptions was introduced
by L.~Landau in 1944~\cite{landau}. However, in particle physics often
the name {\it Landau function} is used generically to refer to all
straggling functions.

The ionization distribution (the distribution of the number of
electron/ion or electron/hole pairs $N_I$ along the particle path
length $x$) $G(x, N_I)$ can be obtained from $F(x, \Delta)$ by using the
relation $N_I=\Delta/W$, assuming that the energy loss is completely
deposited in the material (the sensitive volume of the detector). Also
$G(x, N_I)$ is called a straggling function.

The discussion of straggling functions is generally divided into
two cases: thick absorbers and thin absorbers. In a
{\it thick absorber}, with a thickness sufficient to absorb an amount
of energy comparable to the particle’s initial kinetic energy, the
straggling function can be approximated by a Gaussian
distribution. For {\it thin absorbers}, where only a fraction of the
kinetic energy is lost through ionizing collisions, the straggling
function is always influenced by the possibility of large energy
transfers in single collisions. These add a long tail (the
{\it Landau tail}) to the high energy side, resulting in an asymmetric
form with a mean value significantly higher than the most probable
value.

\subsection{PID using ionization measurements}
\label{sec:dedxmeth}

Measurements of the deposited energy $\Delta$ can be used for
PID. Gaseous or solid state counters\footnote{In a typical particle
  physics experiment the counters used for ionization measurements are
  the same devices that also provide the spatial coordinates for the
  momentum measurement.} provide signals with pulse height $R$
proportional to the number of electrons $N_I$ liberated in the
ionization process along the track length $x$ inside the sensitive
volume, and thus proportional to $\Delta$. To limit deterioration of
the resolution due to energy loss fluctuations, in general $N_R$ pulse
height measurements are performed along the particle track, either in
many consecutive thin detectors or in a large number of samples along
the particle track in the same detector volume. For sufficiently large
$N_R$ the shape of the obtained pulse height distribution approaches
the one of $G(x, N_I)$. The mean value of the distribution is not a
good estimator for the energy deposit $\Delta$, since it is quite
sensitive to the number of counts in the tail, which limits the
resolution. A better estimator is derived from the peak of the
distribution, which is usually approximated by the truncated mean
$\langle R\rangle_a$, which is defined as the average over the $M$
lowest values of the pulse height measurements $R_i$:

\beq
  \label{eq:trunc}
  \langle R\rangle_a = \frac{1}{M}\sum_{i=1}^M R_i \ .
\eeq

\noindent Here $R_i \leq R_{i+1}$ for $i = 1,\ldots,n-1$ and $M$ is
an integer $M=aN_R$. $a$ is a fraction typically between 0.5 and
0.85. Values of $\langle R\rangle_a$ follow an almost perfect Gaussian
distribution with variance $\sigma_E$.

From a theoretical point of view it should be possible to achieve
better results by using  all the information available,~i.e. the shape
of $G(x, N_I)$. A maximum likelihood fit to all $N_R$ ionization
measurements along a track or tabulated energy loss distributions may
be used. However, in practice they show a performance that is not
significantly better~\cite{blumrolandi,bichsel}. Thus usually the
truncated mean method is chosen because of its simplicity.

\begin{figure}[t]
  \centering
  \includegraphics[width=9cm]{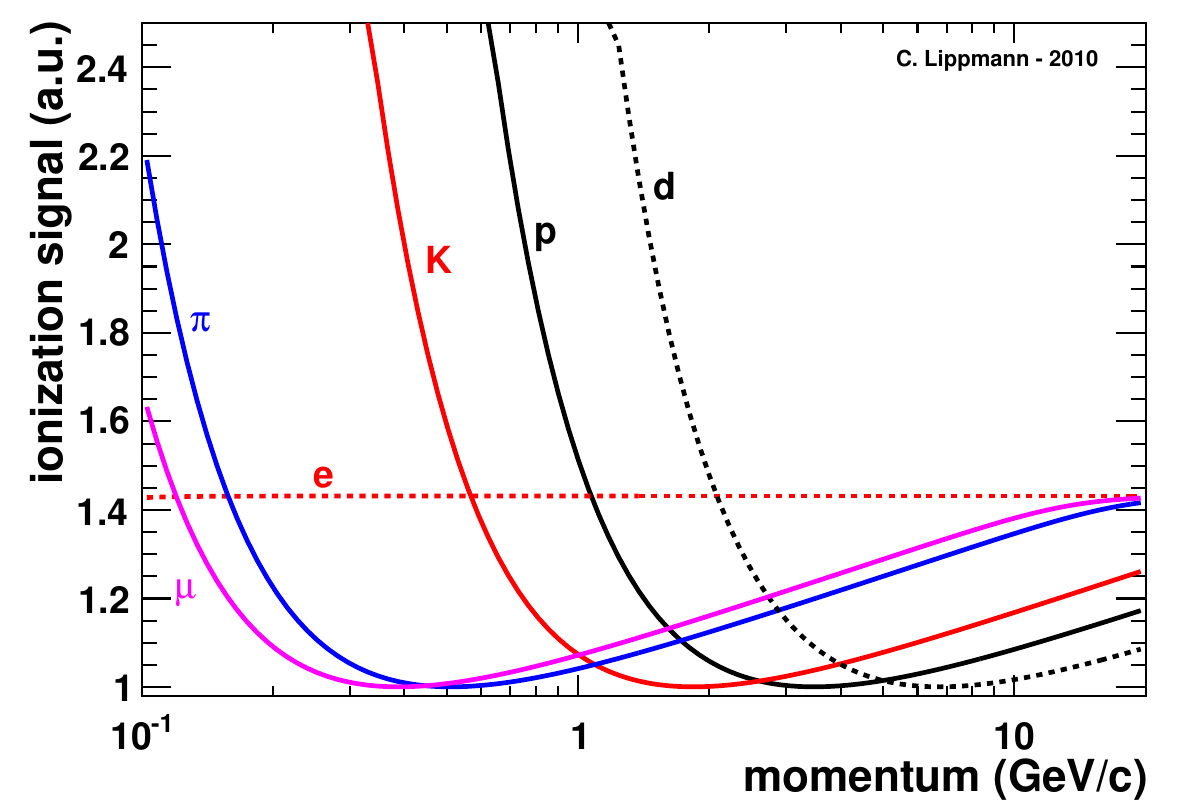}
  \caption{Typical curves of the ionization signal as a function of
    particle momentum for a number of known charged particles. A
    parameterization like the one suggested in Ref.~\cite{blumrolandi}
    was used to calculate the curves.}
  \label{fig:aleph}
\end{figure}

In practice, only relative values of the deposited energy
$\Delta=\Delta (m,\beta)$ are needed to distinguish between different
particle species. Because Eq.~\ref{eq:bb} is not a monotonic function
of $\beta$, it is not possible to combine it with Eq.~\ref{eq:basic} in
order to obtain a form $m=m(\Delta, p)$, where $p$
is the momentum of the particle. However, the PID capability becomes
obvious by plotting ionization curves $\Delta (m, \beta)$ for a number
of charged particles. A description of measured ionization curves
based on five parameters is given in Ref.~\cite{blumrolandi} and can
be used for this purpose. An example is shown in Fig.~\ref{fig:aleph}. 
By simultaneously measuring $p$ and $\Delta$ for any particle of
unknown mass, a point can be drawn on this diagram. The particle is
identified when this point can be associated with only one of the
curves within the measurement errors. Even in areas where the bands
are close, statistical PID methods may be applied.

\subsection{Energy resolution and separation power}
\label{sec:tpcsep}

The resolution in the ionization measurement ({\it energy resolution})
$\sigma_E$ is given by the variance of the Gaussian distribution of
the truncated mean values and is in general proportional to the energy
deposit: $\sigma_E \propto \Delta$. Using Eq.~\ref{eq:sep}, the
separation for two particles $A$ and $B$ with different masses but
the same momentum can be calculated as

\beq
  \label{eq:tpcsep}
  n_{\sigma_E} = \frac{\Delta_A-\Delta_ B}{\langle\sigma_{A,B}\rangle}\ .
\eeq

\begin{figure}[t]
  \centering
  \includegraphics[width=9cm]{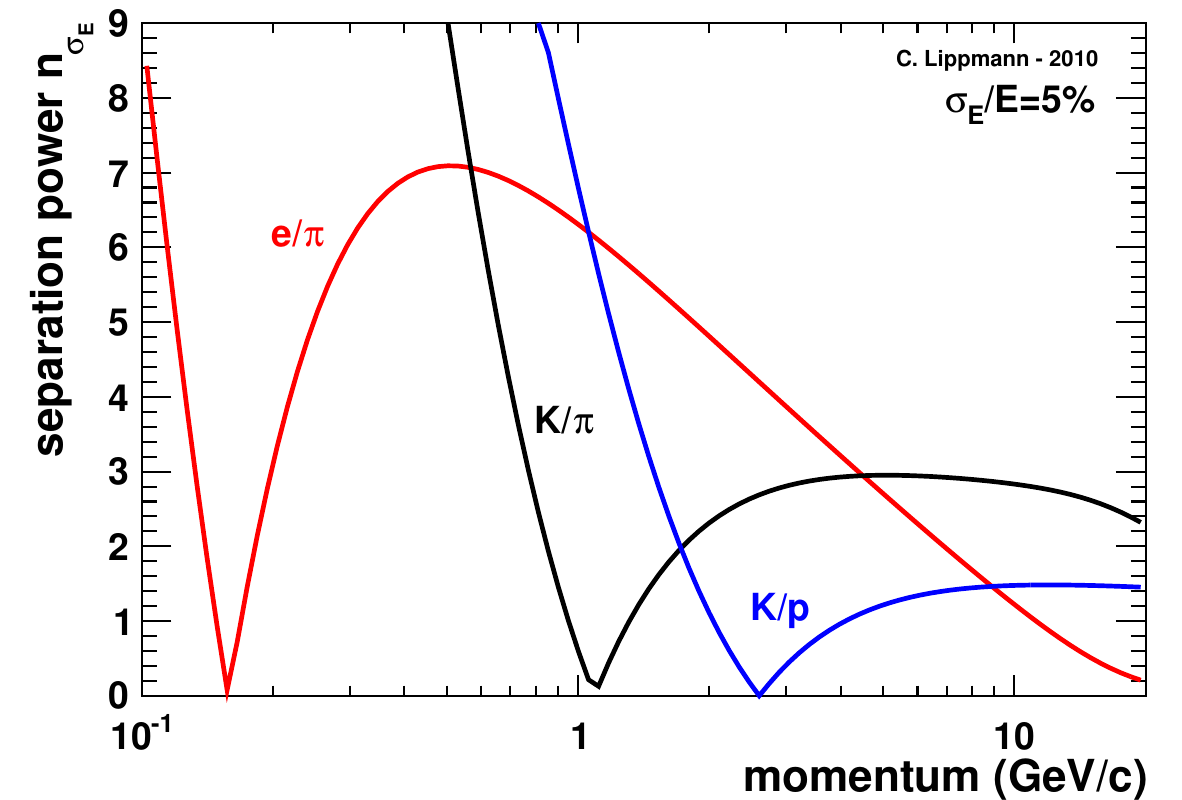}
  \caption{Typical separation power achievable with ionization
    measurements in a gaseous detector. The ionization curves from
    Fig.~\ref{fig:aleph} were used together with an assumed energy
    resolution of 5\%.}
  \label{fig:tpcsep}
\end{figure}

\noindent The average of the two resolutions 
$\langle\sigma_{A,B}\rangle=(\sigma_{E,A}+\sigma_{E,B})/2$ is used.
For the purpose of qualitatively discussing the PID capabilities this
approximation is sufficient. For alternative ways to describe the
resolution of the PID, refer to Ref.~\cite{bichsel}.

In Fig.~\ref{fig:tpcsep} typical particle separations are shown for a
gaseous detector with an energy resolution of 5\%. As expected, hadron
identification works well in the low momentum region. In the minimum
ionization region, where hadrons carry momenta around a few GeV/c and
the ionization curves are very close, it is likely that the method
fails to discriminate the particles. In the  region of the
relativistic rise moderate identification is possible on a statistical basis.

\subsection{Errors affecting the resolution}
\label{sec:dedxlimit}

The operational regime of the particle detector used for the
ionization measurements must be chosen such that the measured pulse
heights $R$ are exactly proportional to the number of electron/ion
(electron/hole) pairs $N_I$ created by the energy deposit
$\Delta$. The same is true for the signal processing
chain. Nevertheless, some effects change the apparent energy deposit
and thus limit the energy resolution of the device. They have to be
considered on top of the fundamental limit that is given by the
statistics of the primary ionization.

\begin{itemize}
\item The ionization signal amplitude is determined by the detector
  (e.g. gas amplification) and by the electronics. An optimal
  performance can be achieved only if an energy calibration is
  carried out in order to determine the absolute gain of each channel
  with high precision.
\item Overlapping tracks must be eliminated with some safety margin.
  This can introduce some limit for a high track density environment like
  heavy-ion collisions or jets, because the number of ionisation
  measurements $N_R$ entering the truncated mean calculation
  (Eq.~\ref{eq:trunc}) can be considerably reduced by this effect.
\item Track independent effects have to be kept under control.
  For a gaseous device these include the gas density (pressure and
  temperature) and gas mixture. Electronics effects like noise and
  baseline shifts have to be taken care of as well.
\item On the level of each track, more effects influence the apparent
  ionization: the detector geometry together with the track
  orientation influence the amount of ionization charge per sampling
  layer and can be corrected for. In a gaseous device, attachment of
  the drifting electrons decreases the ionization signal and requires
  a correction, which depends on the drift length $z$ through a
  constant $b$: $R(z) = R\,e^{-bz}$. A  similar decrease of the signal
  as a function of the drift length can be due to diffusion of the
  drifting primary electron cloud, and can be corrected for as well.
\end{itemize}

For gaseous detectors, these effects are described in detail in
Ref.~\cite{blumrolandi}. An empirical formula exists to estimate the
dependence of the energy resolution on the number of measurements
$N_R$, on the thickness of the sampling layers $x$, and on the gas
pressure $P$~\cite{walenta,nappi}:

\beq
  \label{eq:walenta}
  \sigma_E = 0.41\ N_R^{-0.43} (xP)^{-0.32} \ .
\eeq

\noindent The formula was obtained by fits to measured data and
includes optimizations such as the truncated mean method and the
ones just described. If the ionization measurements along a track were
independent, and if no other error sources like electronics noise
existed, the resolution would scale as $N_R^{-0.5}$. Moreover, the
formula shows that for a fixed total length of a detector $x\,N_R$,
one obtains a better resolution for a finer sampling, provided that
the ionization is sufficient in each sampling layer. A comparison of
the expected resolutions calculated with Eq.~\ref{eq:walenta}  and
measured values is given in Ref.~\cite{ullaland}. Summaries
and examples of performances achieved with gaseous tracking
detectors like Drift, Jet or Time Projection Chambers are given in
Refs.~\cite{blumrolandi,nappi}. Typical values are 4.5 to 7.5\%.
By increasing the gas pressure, the resolution can be significantly
improved,~e.g. for the PEP TPC an energy resolution of below 3\% was
achieved at 8.5\,bar. However, at higher pressure the relativistic
rise is reduced due to the density effect.

\subsection{ALICE Time Projection Chamber}
\label{sec:alicetpc}

\begin{figure}[t]
  \centering
  \includegraphics[width=9cm,angle=90,clip]{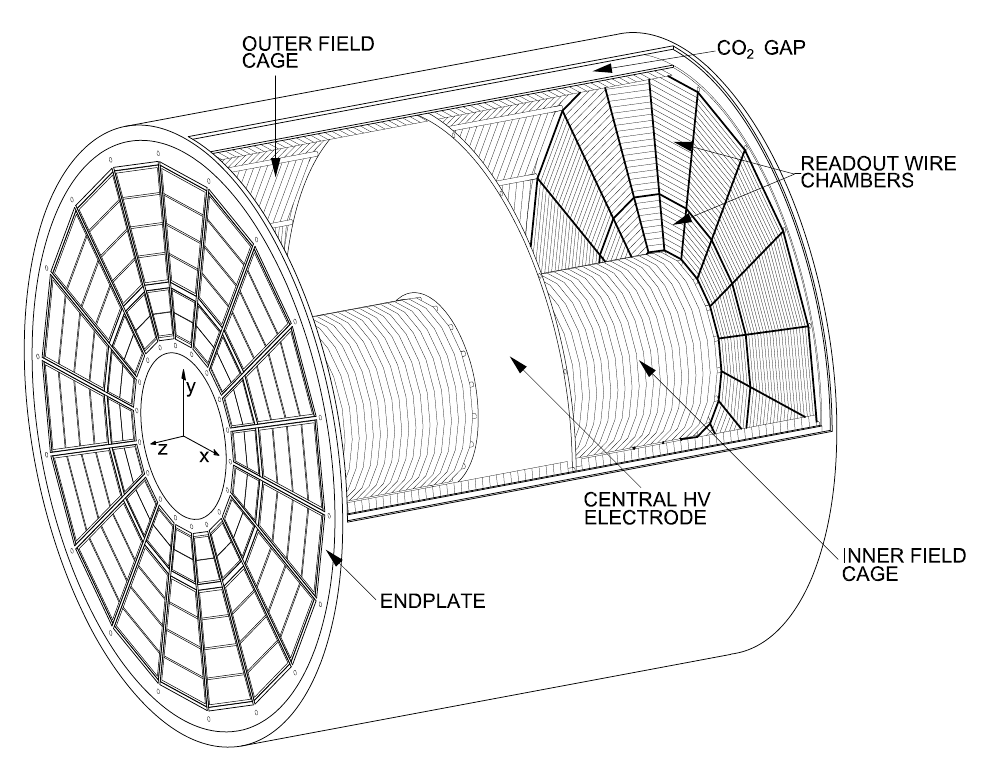}
  \caption{3D view of the TPC field cage~\cite{tpcnim}.  The high
    voltage electrode is located at  the center  of the 5\,m long drift
    volume. The two endplates are divided into 18 sectors holding two
    readout chambers each.}
  \label{fig:tpc}
\end{figure}

The ALICE TPC~\cite{tpctdr,tpcnim} (Fig.~\ref{fig:tpc}) is the largest
gaseous Time Projection Chamber built so far. A TPC is a remarkable
example of a detector performing precise tracking and PID through
measurements of the energy deposited through ionization. Since it is
basically just a large volume filled with a gas, a TPC offers a
maximum of active  volume with a minimum of radiation length. In a
collider experiment like ALICE, the field cage of a TPC is typically
divided into two halves separated by a planar central electrode made
by a thin membrane. The electrons produced by charged particles
crossing the field cage drift towards the two end-caps, where
readout detectors are mounted. In general, a gated wire grid is
installed directly in front of the readout chambers. When the system
is triggered, the gate opens, allowing the passage of the ionization
electrons, which create charge avalanches in the readout
detectors. Closing the gate assures that the ions created in the
avalanche process do not enter the drift volume. Measuring the
signals induced on adjacent readout pads makes possible an
accurate determination of the position by evaluating the center of
gravity. Together with accurate measurements of the
arrival time (relative to some external reference such as the
collision time of the beams) the complete trajectory of all charged
particles traversing the TPC can be determined. Due to the little
amount of dead volume and the easy pattern recognition
(continuous tracks), TPCs are the best tracking devices for high
multiplicity environments,~e.g. heavy-ion experiments.

\subsubsection{Field cage and readout chambers}

The ALICE TPC is a large-volume TPC with overall ``conventional''
lay-out but with nearly all other design parameters pushed towards the
limits. A comprehensive description can be found in
Ref.~\cite{tpcnim}. The TPC field cage is a hollow cylinder
whose axis is aligned with the LHC beam axis and is parallel to
the magnetic field. The active volume has an inner radius of about
85\,cm, an outer radius of about 250\,cm, and an overall length along
the beam direction of 500\,cm. The central electrode is charged to
100\,kV and provides, together with a voltage dividing network at the
surface of the outer and inner cylinder, a precise axial electric
field of 400\,V/cm. To ensure low diffusion of the drifting electrons
and a large ion mobility, Ne was chosen as the main component of the
counting gas. The gas mixture with 10\,\% of CO$_2$ as quencher makes
mandatory a very good temperature homogenisation in the drift volume
($\Delta T < 0.1$\,K), in order to guarantee a homogeneous drift
velocity. It is operated at atmospheric pressure.

Wire chambers with cathode pads are used for the readout of the
TPC. An optimization of the design for the expected
high track density environment implied rather small pad sizes
($4\times 7.5$\,mm$^2$ in the innermost region). As a consequence,
the number of pads is large ($\sim$560\,000; for comparison: the
ALEPH TPC had 41\,000 pads), and the readout chambers have to be
operated at rather high gas gains near $10^4$.

\subsubsection{Front-end electronics}

Also for the design of the front-end electronics the key input was
the operation in a high track density environment~\cite{tpcnim}. A
signal occupancy as high as 50\% can be expected for central Pb--Pb
collisions. Due to the large raw data volume (750\,MB/event) zero
suppression is implemented in the front-end electronics, thus reducing
the event sizes to about 300\,kB for p--p collisions (no overlying
events) and to about 12\,MB for Pb--Pb collisions. Signal tail
cancellation and baseline correction are performed before the zero
suppression in order to preserve the full resolution on the signal
features (charge and arrival time). A large dynamic range ensures that
the ionization signal of particles can be determined with precision
from very low up to high momenta.

\subsubsection{PID performance and outlook}

A very good energy calibration is fundamental for an optimal PID
performance (see Sect.~\ref{sec:dedxlimit}). In the ALICE TPC, the
amplitude of the ionization signal is determined by the gas
amplification and the electronics. The absolute gain of each channel
is obtained with high precision by periodically releasing
radioactive krypton ($^{83}_{36}$Kr) into the TPC gas system and
measuring the well known decay pattern (for more details see
~\cite{tpcnim}).

\begin{figure}[t]
  \centering
  \includegraphics[width=12cm]{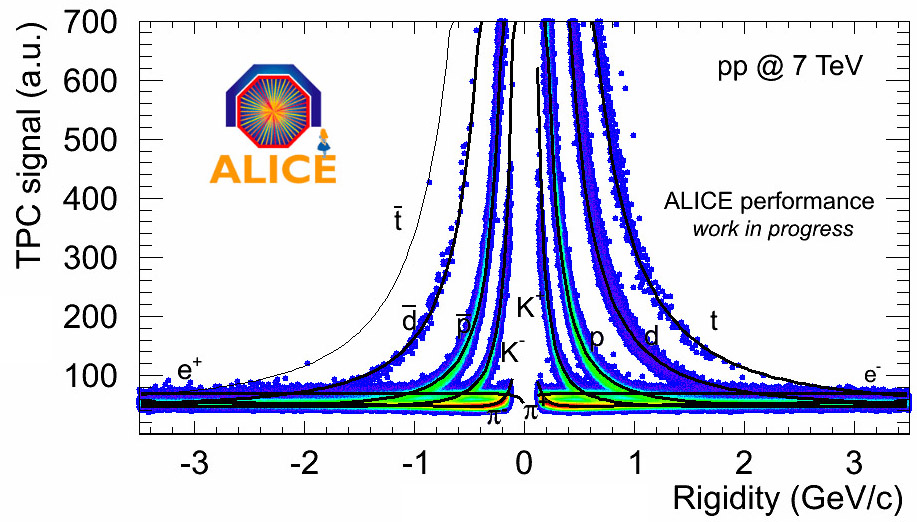}
  \caption{Measured ionization signals of charged particles as a
    function of the track rigidity (particle momentum divided by
    charge number). $11\times 10^{6}$ events from a data sample
    recorded with $\sqrt{s}=7$\,TeV p--p collisions provided by the
    LHC were analysed. The lines correspond to a parameterization of
    the Bethe-Bloch curve as described in
    Ref.~\cite{blumrolandi}. Also heavier nuclei like deuterons and
    tritium and their anti-particles are found.}
  \label{fig:tpcperf}
\end{figure}
 
\begin{figure}[t]
  \centering
  \includegraphics[width=10cm]{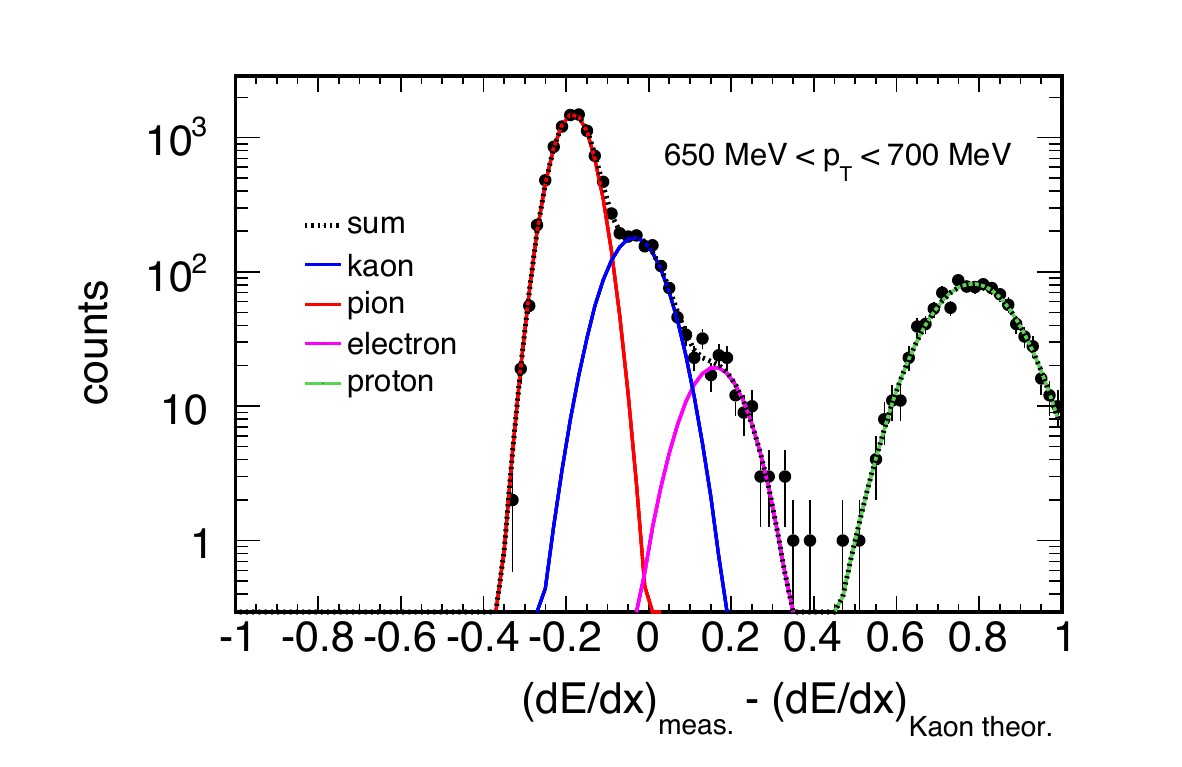}
  \caption{Distribution of the difference between the measured
    ionization signals and the one expected for Kaons for a momentum
    slice of 50\,MeV/c width. The lines are fits, indicating that a
    sum of four Gaussians represents well the data. The peak centered
    at zero reflects the abundance of kaons, the other peaks represent
    other particle species.}
  \label{fig:tpcgaus}
\end{figure}

Inside the TPC, the ionization strength of all tracks is sampled on up
to 159 pad rows. The energy resolution is 5\% for tracks with the
maximum number of signals. Fig.~\ref{fig:tpcperf} shows the ionization
signals from $11\times 10^{6}$ events from a data sample recorded with
$\sqrt{s}=7$\,TeV p--p collisions provided by the LHC. Clearly the
different characteristic bands for various particles including
deuterons and tritium are visible for particles and anti-particles,
respectively. In order to achieve PID in a certain momentum region,
histograms are filled and fitted with multiple Gaussians. An example
is shown in Fig.~\ref{fig:tpcgaus}. For a given ionization signal in
this momentum region one can derive the particle yields, which
represent the probabilities for the particle to be a kaon, proton,
pion or electron.


The ALICE TPC is in nearly continuous data taking mode since more than
one year, providing tracking and PID information while the LHC
delivers p--p collisions at $\sqrt{s}=7$\,TeV. In November 2010 the
first Pb--Pb collisions at $\sqrt{s_{NN}}=2.76$\,TeV were
recorded. The energy resolution of the TPC in these heavy-ion
collisions was found to degrade slightly to an average of about
5.3\%. This effect is expected due to overlapping signals from
neighboring tracks. The energy resolution for the highest multiplicity
events falls off by 20\% with respect to low multiplicities. Part of the
increase is also explained by baseline fluctuations in the electronics
due to large hit densities in single channels. These fluctuations can
in the future be minimised using the signal tail cancellation and
baseline correction features available in the TPC front-end electronics.
In the course of the year 2011, higher luminosities will be reached
and space charge effects and gating efficiency will have to be
accurately evaluated to avoid negative effects on the tracking and
ionization sampling.

\subsection{Silicon detectors at LHC}
\label{sec:dedxother}

At the LHC, the signal amplitude information of the silicon strip
(ALICE, ATLAS and CMS) and drift (ALICE) detectors of the tracking
systems make possible hadron ID via ionization measurements. However,
since in solid media due to a stronger density effect the Fermi
plateau is only slightly above the minimum, good $\pi$/K (p/K)
separation is achievable only up to 450\,MeV/c (up to
1\,GeV/c). Again, the energy loss is estimated as a truncated mean in
order to minimize the influence of Landau fluctuations. In ALICE, the
resolution of the measurement in two layers each of silicon strip and
drift detectors is about 11\%. In the CMS tracking system, the
signal amplitudes from ten layers of silicon strip detectors may be
used, but a resolution was not known to the author at the time of
writing this review.

\subsection{Developments for future TPCs}
\label{sec:ilctpc}

A TPC at a future linear collider will very likely be read out by
micro-pattern devices like GEMs~\cite{gem} or
MicroMegas~\cite{micromegas}. Their advantages are the high rate
capability and a low ion feedback. Large TPC protopypes with different
readout options were built tested~\cite{ilctpc}. The conventional readout 
of the micro-pattern device, with mm-sized cathode pads, may actually
be replaced by a digital device with pixel sizes of the order
55\,$\mu$m. A family of ASIC chips is particularly
suited for this purpose: MEDIPIX~\cite{medipix} and its successor
TIMEPIX~\cite{timepix}, which records also the time. This kind of setup
has been shown to detect tracks of minimum ionizing charged particles
(MIPs) with excellent single-electron efficiency and unprecedented
spatial resolution. In this way, the  {\it cluster counting}  method
may be used to assess the energy deposit through ionization: rather
than measuring the deposited charge, the number of primary collision
clusters are counted. This avoids the problems related to the
fluctuation of the energy transfer in single collisions (see
Sects.~\ref{sec:straggling} and \ref{sec:dedxmeth}) and should provide
the ultimate resolution of $\sim$2\%~\cite{cluster}. Concerning
the implementation in a large detector like a TPC, the
challenges will include equipping a large active area with the
pixelized readout and handling huge amounts of readout channels.

\section{Time-of-flight}
\label{sec:tof}

Time-of-flight (TOF) measurements yield the velocity of a charged
particle by measuring the particle flight time $t$ over a given
distance along the track trajectory $L$. The particle velocity
$\beta=v/c=L/tc$ depends on its mass $m$ and momentum $p$
through

\beq
  \label{eq:tof1}
  \beta = \frac{1}{\sqrt{\left(\frac{mc}{p}\right)^2+1}} \ .
\eeq
 
\noindent Thus, one can calculate the mass $m$ from measurements
of $L$, $t$ and $p$:

\beq
  \label{eq:tof}
  m = \frac{p}{c}\sqrt{\frac{c^2t^2}{L^2}-1} \ .
\eeq

\subsection{Time resolution and separation power}
\label{sec:tofsep}

If two particles with masses $m_A$ and $m_B$, respecively, carry the
same momentum, their flight time difference can be calculated as

\beq
  \label{eq:tof2}
  |t_A - t_B| = \frac{L}{c} \left| \sqrt{1+\left( \frac{m_A c}{p} \right)^2}
    - \sqrt{1+\left( \frac{m_B c}{p} \right)^2} \ \right| \ .
\eeq

\noindent With $p\gg mc$ the approximation
$\sqrt{1+(mc/p)^2} \approx 1+(mc)^2/2p^2$ can be used, and with
Eq. \ref{eq:sep} the separation power becomes

\beq
  \label{eq:tofsep}
  n_{\sigma_{TOF}} = \frac{|t_A - t_B|}{\sigma_{TOF}} =
  \frac{Lc}{2p^2\sigma_{TOF}} |m_A^2 - m_B^2| \ . 
\eeq
 
\begin{figure}[t]
  \centering
  \includegraphics[width=9cm]{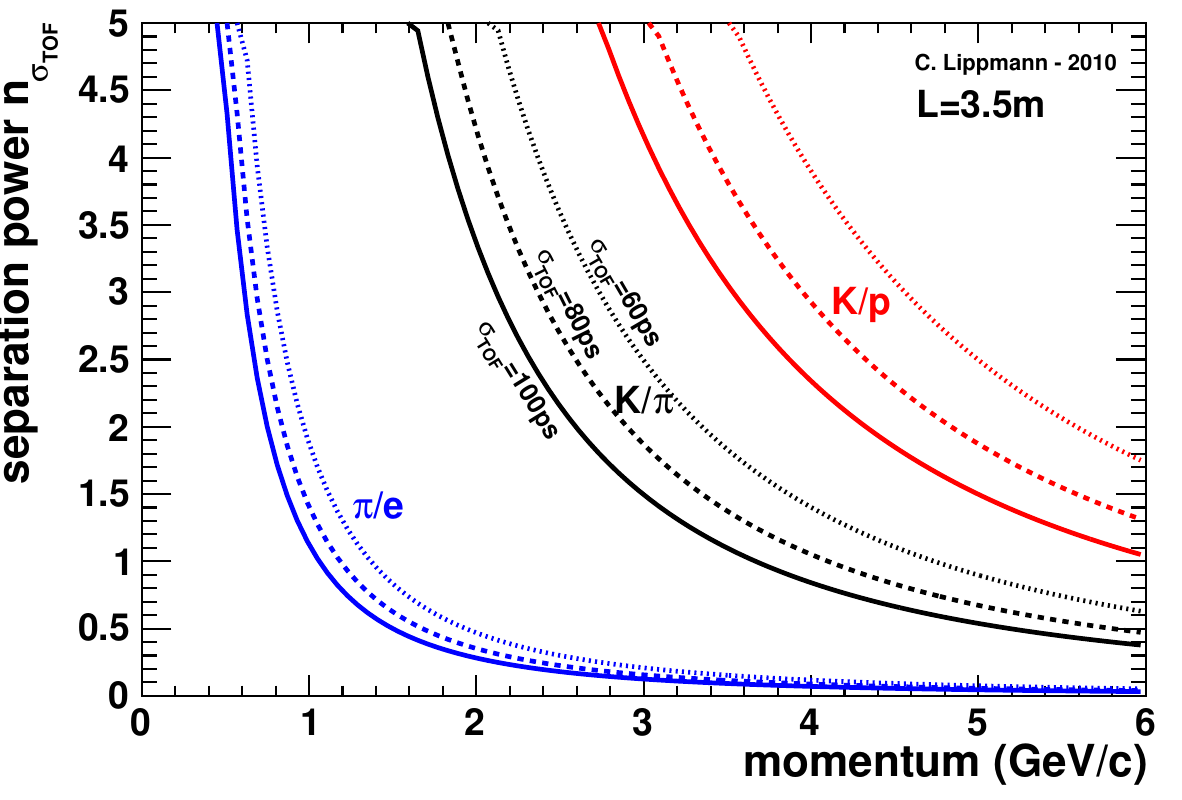}
  \caption{Particle separation with TOF measurements for
    three different system time resolutions ($\sigma_{TOF}=60$, 80 and
    100\,ps) and for a track length $L=3.5$\,m. Infinitely good
    precisions on momentum and track length measurements are
    assumed.}
  \label{fig:tofsep}
\end{figure}

\noindent Here $\sigma_{TOF}$ is the resolution of the TOF
measurement. Misidentification of particles occurs at higher momenta,
where the time difference $|t_A - t_B|$ becomes comparable to
$\sigma_{TOF}$. Assuming  a time resolution of 100\,ps (60\,ps)
and requiring a separation of $n_{\sigma_{TOF}}=3$, the upper limits
for the momentum are 2.1\,GeV/c (2.7\,GeV/c) for K/$\pi$
separation and 3.5\,GeV/c (4.5\,GeV/c) for K/p separation (see
Fig. \ref{fig:tofsep}). A lower momentum threshold is defined by the
curvature of the tracks in the magnetic field. Assuming a path length
$L=3.5$\,m and a magnetic field of $0.5$\,T, only particles with a
momentum larger than about $300$\,MeV/c reach the TOF wall.

\subsection{Errors affecting the resolution}
\label{sec:toflimit}

The mass resolution of a TOF measurement is given by:


\beq
  \label{eq:toferror}
  \left(\frac{\dd m}{m}\right)^2 = \left(\frac{\dd p}{p}\right)^2 +
  \left(\gamma^2 \frac{\dd t}{t}\right)^2 +
  \left(\gamma^2 \frac{\dd L}{L}\right)^2 \ .
\eeq

\noindent Since in most cases $\gamma \gg 1$, the mass resolution is
predominantly affected by the accuracies in the time and length
measurements, rather than by the accuracy of the momentum
determination. In general the track length $L$ and the momentum $p$
can be measured with rather good accuracies in the tracking system
($\sigma_p/p\approx 1$\%, $\sigma_L/L\approx 10^{-3}$). The
time-of-flight $t$ is measured with a certain accuracy $\sigma_{TOF}$
by a dedicated detector system. A TOF measurement always consists of
two time measurements: $t=t_1-t_0$.  The event start time $t_0$ is
measured with accuracy $\sigma_{t_0}$. The stop time $t_1>t_0$ is
measured with accuracy $\sigma_{t_1}$ by the actual TOF detector,
usually a large area detector system some distance away from the
interaction region. The overall TOF resolution is then given by
$\sigma_{TOF}^2=\sigma_{t_0}^2+\sigma_{t_1}^2$.

\subsection{Resistive Plate Chambers for time-of-flight measurements}
\label{sec:rpctof}

In order to optimise a TOF system the resolution $\sigma_{TOF}$ has to
be improved. In current experiments for
subnuclear research, detectors using scintillation  (the property of
luminescence when excited by ionizing radiation) are very commonly
used due to their good time resolution. However, as the experiments
have gradually increased in size, and since the readout of
scintillator detectors is expensive, the cost per area became an
important issue. The Resistive Plate Chamber (RPC) was found to
provide a simpler technology at a much lower price. RPCs can be seen
as the successor of the metallic Parallel Plate Avalanche Counters,
overcoming their disadvantage of being very vulnerable to the
destructive energy released by possible discharges or sparks. The
electrode plates of RPCs are made from a resistive material, which
effectively restricts excessive signals or discharges to a well
localised area of the detector. As a side effect, all charges cause a
local drop of the electric field in the gas gap and in this spot the
detector becomes insensitive to further  traversing particles for a
time of the order of the relaxation time
$\tau = \rho\,\eps_0\eps_r$, where $\eps_0$ is the dielectric constant
and $\rho$ and $\eps_r$ are the volume resistivity and the relative
permittivity of the resistive material. Even though the
remaining counter area is still sensitive to particles, a limit to the
overall rate capability is introduced through this effect. In general,
accurate time measurements are feasible up to charged particle rates
of a few kHz/cm$^2$, which is acceptable for many purposes (compare
to Sect.~\ref{sec:tofnew}).

The RPC design was improved by introducing the concept of
multiple gaps (Multi gap Resistive Plate Chamber,
MRPC~\cite{multigap}), resulting in an increase of detection
efficiency. Reducing the individual gap sizes~\cite{thingap}, while at
the same time increasing the electric field, resulted in what is called
the {\it Timing RPC}, a low cost detector with a time resolution
comparable to that of scintillators. The space resolution can be
chosen by different sizes of the readout strips.

\subsection{Detector physics of Resisitive Plate Chambers}
\label{sec:rpc}
 
The physics of avalanches in RPCs is very complex, even though the
simple geometry of the device would not suggest so. During the
development of large area RPC systems for the LHC experiments, the
detector physics of RPCs was therefore studied in
detail~\cite{rpctheory,spacecharge,rpcrate,aielli,abrescia,mangiarotti1,mangiarotti2,blanco}.
The good detection efficiency of (single gap) RPCs is explained by the
very large gas gain. A very strong space charge effect reduces the
avalanche charges by many orders of magnitude. Detailed simulations
show that the space charge field inside the avalanches actually
reaches the same magnitude as the applied electric field. 

The very good time resolution of RPCs is due to the strong uniform
electric field, which provokes the avalanche process immediately after
primary ionization is deposited in the gas volume. As a consequence,
the intrinsic detector time resolution (no electronics) is determined
by the avalanche statistics, and can be estimated using a simple
formula (for one gas gap):

\beq
  \sigma_{Intr} \approx \frac{1.28}{(\alpha_g-\eta_g)v_D}\ .
  \label{analTR}
\eeq

\noindent Here $\alpha_g$ and $\eta_g$ are the Townsend and attachment
coefficients of the gas mixture and $v_D$ is the the drift velocity of
electrons. Typical operational parameters~\cite{rpctheory}
for Timing RPCs are $\alpha_g = 123$\,mm$^{-1}$,
$\eta_g=10.5$\,mm$^{-1}$ and $v_D=210$\,$\mu$m\,ns$^{-1}$. For
these values an intrinsic time resolution of $\sigma_{Intr}=50$\,ps is
obtained for a single gap. By increasing the number of gaps $n$, the
resolution can be improved, but for small values of $n$ the
improvement is not following a simple $1/\sqrt{n}$ scaling, because it
is dominated by the gap which has the largest
signal~\cite{rpctheory}. For $n=10$ gaps a resolution of
$\sigma_{Intr}=20$\,ps is feasible.

The drop of the electric field in the gas gap at high particle rates affects
efficiency and time resolution. It may be calculated in a simple way
by assuming the particle flux to be a DC current that causes a voltage
drop when it passes through the resistive plate. For a single gap of
width $b$ the average field reduction is:

\beq
  \label{eq:rpcrate}
  \langle \Delta E \rangle = \rho \frac{a}{b} \Phi Q\ ,
\eeq

\noindent where $\rho$ and $a$ are the volume resistivity and thickness
of the resistive electrode material and $\Phi$ is the particle flux per area.
$Q$ is the avalanche charge, which of course depends itself on the
field reduction, and has thus to be found in an iterative
procedure. An approximate analytic expression for the field
fluctuations around $\langle \Delta E \rangle$ is given in
Ref.~\cite{rpcrate}.

\subsection{ALICE time-of-flight detector}
\label{sec:alicetof}

\begin{figure}[t]
  \centering
  \includegraphics[width=10cm]{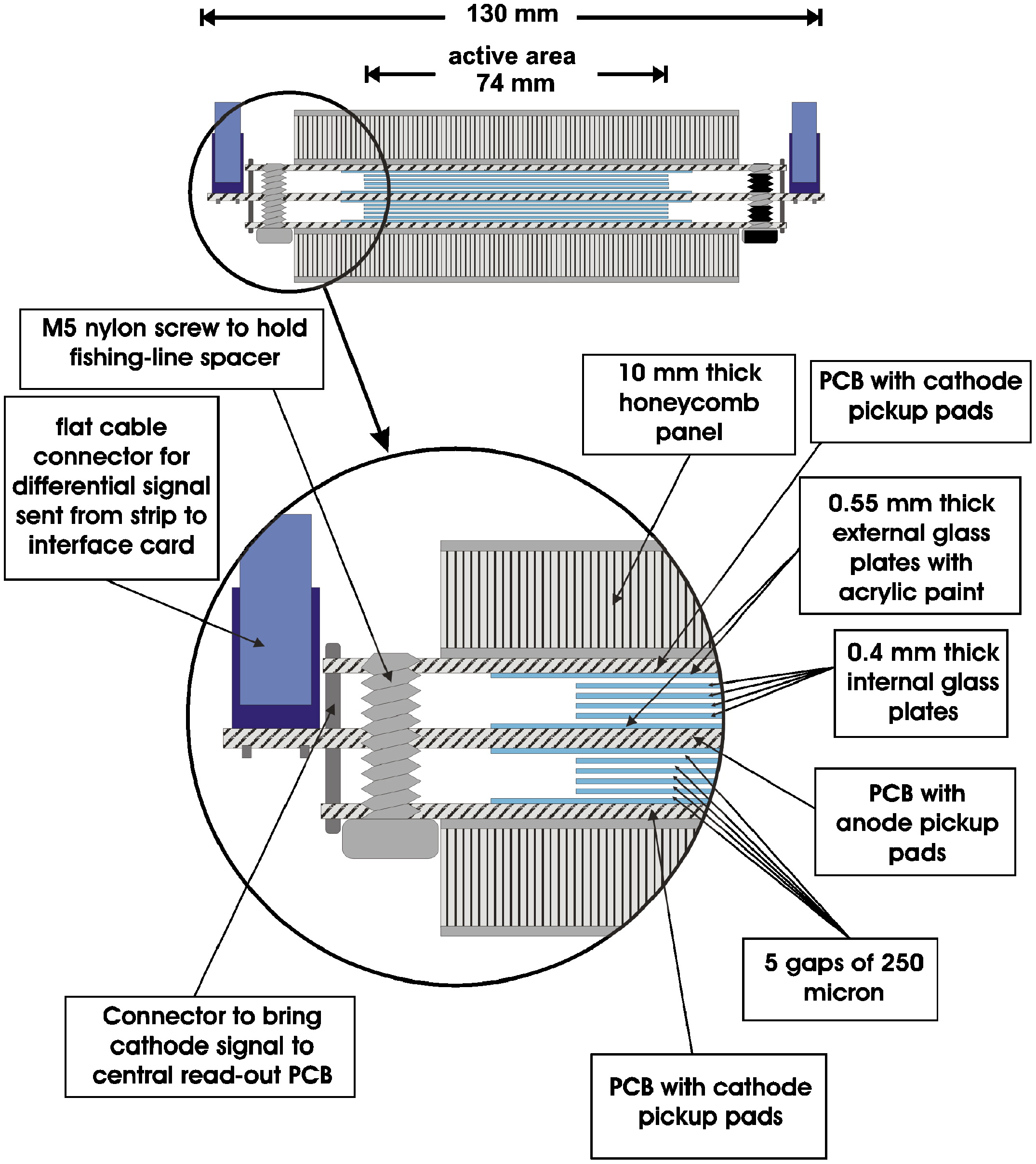}
  \caption{Cross-section of an ALICE MRPC~\cite{alice}.}
  \label{fig:tof}
\end{figure}
 
In specific momentum ranges, the PID signals from the ALICE TPC can
not be used to separate kaons, pions and protons, because the
ionization curves are very close or overlap (see also
Fig.~\ref{fig:summary}). The NA49 and STAR experiments have
shown~\cite{na49,starpid} how the hadron identification capability may
be extended by combining the PID informations from TPC and
TOF. Moreover, a reliable identification of electrons is possible up
to a few GeV/c in this way, since the hadrons (which for a given
momentum are slower) can be eliminated with a cut on the particle
velocity.

The ALICE TOF detector~\cite{toftdr} is based on MRPCs. 91 MRPCs each
are assembled into gas tight modules in 18 azimuthal sectors
(supermodules) surrounding the ALICE TPC and TRD (see
Fig.~\ref{fig:alice}). The supermodules have about 9\,m length,
covering the pseudo-rapidity interval $|\eta|<0.9$. More than 150\,000
readout pads with size $2.5\times 3.5$\,cm$^2$ cover an area of about
150\,m$^2$. The gas mixture in the ten 250\,$\mu$m thin gas gaps is
90\% C$_2$H$_2$ F$_4$, 5\% C$_4$H$_{10}$, 5\%  SF$_6$. A cross-section
through one MRPC is shown in Fig.~\ref{fig:tof}.

\subsubsection{Time resolution}

The design goal for the ALICE TOF was to reach a global
TOF resolution of $\sigma_{TOF}\approx 100$\,ps. The following
contributions have to be considered:

\beq
  \label{eq:tofressum}
  \sigma_{TOF}^2 \ = \ \sigma_{t_0}^2+\sigma_{t_1}^2 \ = \
  \sigma_{t_0}^2+\sigma_{Intr}^2+\sigma_{Elec}^2
  +\sigma_{Clock}^2+\sigma_{Cal}^2 \ .
\eeq
 
\noindent Here $\sigma_{t_0}$ is the uncertainty in the absolute time
of the collision, $\sigma_{Intr}$ is the intrinsic time resolution of the detector
technology ($\sim$20\,ps) and $\sigma_{Elec}$ combines the intrinsic
time jitter of the amplification electronics ($\sim$20\,ps) and of the
time-to-digital conversion ($\sim$30\,ps). $\sigma_{Clock}$ is an
uncertainty arising from the distribution of the digital clock from
the LHC to the experiment ($\sim$15\,ps) and through the electronics
chain ($\sim$10\,ps). Finally, a contribution $\sigma_{Cal}$ includes
all effects that can be parametrised and that are thus accessible to
calibration methods in order to be minimised. These are mainly

\begin{enumerate}
\item the relative timing of the different channels given by cable
  lengths ({\it time offset}),
\item the influence of the finite rise time of the ampliflying
  electronics\footnote{In the ALICE TOF front-end electronics the
    channel-to-channel time slewing corrections are measured by a
    Time-Over-Threshold circuit, which provides a time width
    approximately proportional to the signal charge.}
  ({\it time slewing}) and
\item the influence of the impact position of the particle through
  signal propagation delays on the pad ({\it time walk}).
\end{enumerate}

\begin{figure}[t]
  \centering
  \includegraphics[width=10cm]{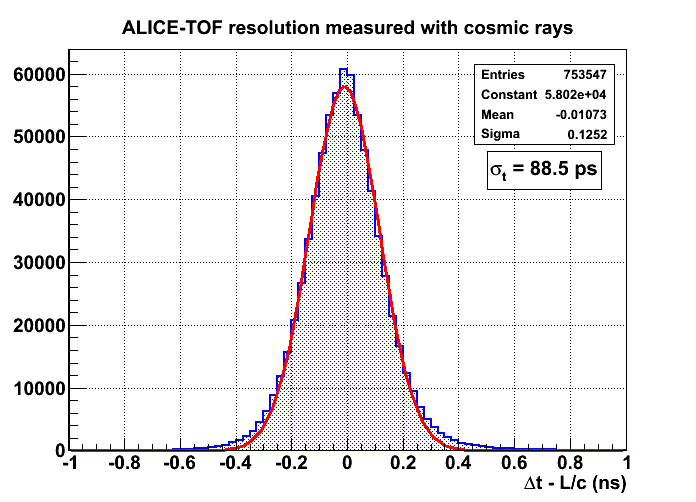}
  \caption{TOF measurements with a cosmic data sample from
    the year 2009. Since the start and stop time were both measured
    with the ALICE TOF, the resolution of a single time measurement is
    obtained from the histogram by dividing its width by
    $\sqrt{2}$. The resulting resolution of 88.5\,ps corresponds to
    $\sigma_{t_1}$ from Eq.~\ref{eq:tofressum}.}
  \label{fig:tofsig}
\end{figure}

\noindent It is forseen that the contribution $\sigma_{Cal}$ can be
kept below 30\,ps~\cite{ppr2}. Already with the limited statistics
available from cosmics data samples before the start of the LHC it
was possible to reach a value of $\sigma_{Cal}=70$\,ps. The
resolution $\sigma_{t_1}$ with this ``intermediate calibration''
(compare to Tab.~\ref{tab:tof}) is shown in Fig.~\ref{fig:tofsig}.

The absolute time of the collision $t_0$ is fluctuating with respect to
the nominal time signal from the LHC according to a Gaussian
distribution with a $\sigma$ of about 140\,ps. This is due to the
finite length of the colliding bunches of particles. In ALICE, $t_0$ is
precisely measured by a dedicated detector\footnote{Alternatively,
  the information from the TOF detector itself can be used to determine
  $t_0$ in events with at least three tracks with an associated TOF
  signal by means of a combinatorial algorithm.} (T0)~\cite{alice},
which reduces this uncertainty to $\sigma_{t_0}=50$\,ps.

\begin{table}[ht]\footnotesize
  \caption{Overall TOF resolution together with the two main contributions
    $\sigma_{t_0}$ and $\sigma_{Cal}$.}
  \label{tab:tof}
  \begin{center}
    \begin{tabular}{|c||c|c|c|c|c|c|}
      \hline
      $\sigma_{TOF}$ & $\sigma_{t_0}$ & $\sigma_{Cal}$ & Remark \\
      \hline
      \hline
      100 & 50 & 70 & proton collisions, intermediate
      calibration \\
      \hline
      80  & 50 & 30 & proton collisions, optimal calibration \\
      \hline
      65  & 10 & 30 &  Pb--Pb collisions, optimal calibration \\
      \hline
 \end{tabular}
  \end{center}
  \vspace{1mm}
\end{table}

With the much larger statistics that is available from data samples
with collisions provided by the LHC, the work now focuses on reducing
the contribution $\sigma_{Cal}$. A further improvement can be achieved
for high multiplicity Pb--Pb events, where the accuracy of the event
start time measurement is expected to improve to
$\sigma_{t_0}=10$\,ps. In that case an overall system resolution
$\sigma_{TOF}=65$\,ps is possible (compare to Tab.~\ref{tab:tof}).

\subsubsection{PID performance and outlook}

\begin{figure}[p]
  \centering
  \includegraphics[width=12cm]{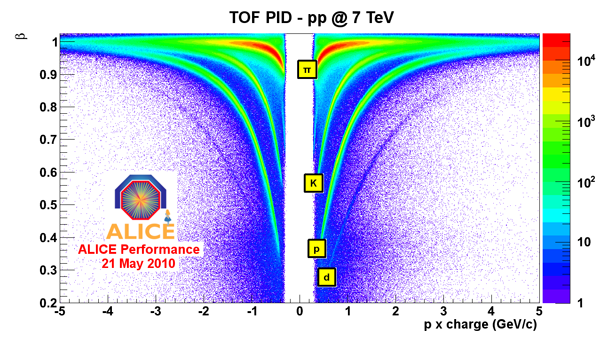}
  \caption{Velocity $\beta=v/c$ as measured with the ALICE TOF
    detector as a function of the particle momentum $p$ multiplied
    with the particle charge number $Z$ for a data sample taken
    with $\sqrt{s}=7$\,TeV collisions provided by the LHC in the year
    2010. No data is available for momenta $\lesssim$300\,MeV/c, since
    these particles do not reach the detector due to the curvature of
    their tracks in the magnetic field.}
  \label{fig:tofpid}
\end{figure} 

\begin{figure}[p]
  \centering
  \includegraphics[width=10cm]{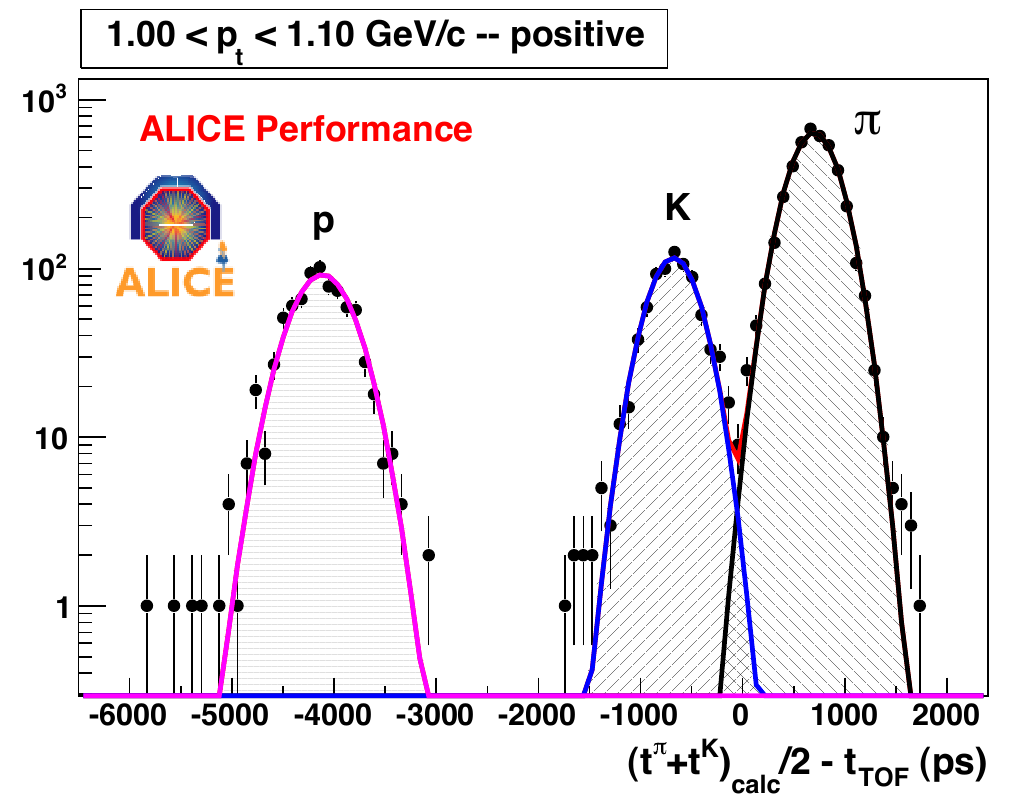}
  \caption{Histogram of the TOF signal for a given momentum window of
    100\,MeV/c width. The data sample was recorded with
    $\sqrt{s}=900$\,GeV p--p collisions provided by the LHC. The
    lines are fits to the data, indicating that a sum of three
    Gaussians well represents the data.}
  \label{fig:tofgaus}
\end{figure}
 
\begin{figure}[t]
  \centering
  \includegraphics[width=10cm]{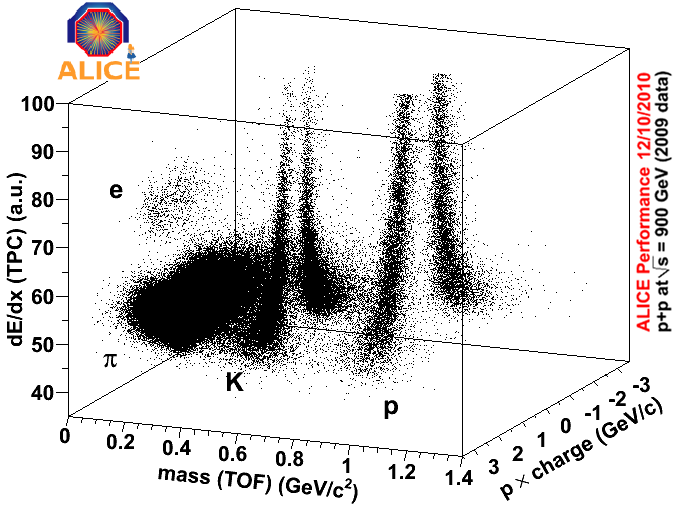}
  \caption{Demonstration of ALICE's PID capabilities by combining the
    ionization measurements in the TPC and the mass calculated using
    the TOF signal. Electrons, pions, kaons and protons are clearly
    visible in a wide momentum range. The data sample was recorded with
    $\sqrt{s}=900$\,GeV p--p collisions provided by the LHC.}
  \label{fig:tof3D}
\end{figure}

Fig.~\ref{fig:tofpid} shows the correlation between the track momentum
measured by the ALICE tracking detectors and the velocity $\beta$ as
measured by the TOF system. The data were recorded during the first
$\sqrt{s}=7$\,TeV p--p collisions in 2010 and clearly show two times
four bands corresponding to four particle species and their corresponding
anti-particles. Data outside these bands indicates errors in the
association of tracks to TOF PID signals. A particle is identified
when the corresponding point in the diagram of Fig.~\ref{fig:tofpid}
can be associated with only one theoretical curve (Eq.~\ref{eq:tof1})
within the measurement errors. For certain momentum regions histograms
are filled and fitted with multiple Gaussians, similar to the
procedure described in Sect.~\ref{sec:alicetpc}. An example is shown
in Fig.~\ref{fig:tofgaus}. For a given TOF measurement one can derive
the particle yields, which represent the probabilities for the
particle to be a kaon, proton, pion or electron. The power of
combining TOF and ionization measurements for PID is demonstrated for
in Fig.~\ref{fig:tof3D}.

\subsection{Other TOF detectors}
\label{sec:tofother}

RPCs have found applications in many particle physics experiments
in many variations, as can be seen in the proceedings of the RPC
workshop series~\cite{tof1,tof2,tof3,tof4,tof5,tof6}. The STAR
experiment has installed a large area (64\,m$^2$) TOF system similar
to the ALICE TOF in their barrel, surrounding the
TPC~\cite{startof1,startof2}. The modules have six gas gaps of
220\,$\mu$m width and are filled with a gas mixture of 95\%
C$_2$H$_2$F$_4$ and 5\% i-C$_4$H$_{10}$. The STAR experiment
also plans to install a Muon Telescope Detector (MTD) based on Timing
RPCs to identify muons~\cite{starmtd}. The FOPI experiment has
upgraded their TOF system in 2007 by installing Multistrip Multigap
Resistive Plate Chambers (MMRPCs) with 8 gaps of
220\,$\mu$m~\cite{fopitof}. For the HADES experiment a similar
upgrade of the inner TOF wall is planned for the near
future~\cite{hadestof}.

\subsection{Developments for future TOF systems}
\label{sec:tofnew}

Driven by the environment in which TOF systems are planned to be
operated in future experiments, a number of R\&D projects currently
address the rate limitation of Timing RPCs. The requirements for the TOF
wall of the CBM experiment, for example, include a rate capability up
to 25\,kHz/cm$^2$ with a time resolution of 80\,ps on a large system
(120m$^2$)~\cite{cbmtof1}. The effective drop of the operating voltage
at larger rates (see Sect.~\ref{sec:rpctof}) can be reduced either by lower
electrode resistivity or by thinner electrode plates. Different electrode
materials are being explored, including low-resistive phosphate and
silicate glass~\cite{cbmtof2,rpcrate1}. With electrodes made from
ceramics the practical feasibility of accurate timing measurements
with RPCs at rates up to 500\,kHz/cm$^2$ was established in test
beams~\cite{rpcceramics,rpcceramics2}.

\begin{figure}[t]
  \centering
  \includegraphics[width=7cm]{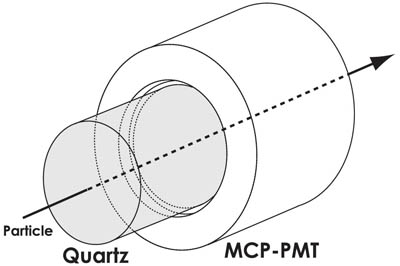}
  \caption{Schematic drawing of a TOF counter based on a
    MCP-PMT~\cite{richtof2}. The MCP-PMT actually detects
    Cherenkov photons emitted in the quartz glass entrance window and
    has a very good time resolution.}
  \label{fig:mcptof}
\end{figure}

TOF measurements with charged particles can also be carried out using
Micro-Channel Plate (MCP) Photo Multiplier Tubes (PMTs, see
Fig.~\ref{fig:mcptof})), which allow time measurements with a few ps
precision. The MCP-PMT actually detects Cherenkov photons emitted in
the entrance window of the PMT (for details on Cherenkov radiation see
Sect.~\ref{sec:rich})~\cite{richtof1}. Time resolutions of
$\sigma_{t_1} \approx 6$\,ps (with an intrinsic resolution of the
detector of $\sigma_{Intr}<5$\,ps) were observed in test
beams~\cite{richtof2,richtof3}. At this scale, the resolution of the
measurement of the event start time $\sigma_{t_0}$ becomes dominant
and must be minimised as well. It remains to be shown that this
technology can be implemented on large surfaces at an affordable
price and can thus compete with MRPCs.

\section{Cherenkov imaging}
\label{sec:rich}

In 1958 P.~A.~\u{C}erenkov, I.~Y.~Tamm and I.~M.~Frank were awarded
with the Nobel Prize in physics for the discovery and interpretation
of the Cherenkov effect. Cherenkov radiation is a shock wave resulting
from a charged particle moving through a material faster than the
velocity of light in the material. The Cherenkov radiation propagates
with a characteristic angle with respect to the particle track
$\Theta_C$, that depends on the particle velocity:

\beq
  \label{eq:cangle}
  \cos(\Theta_C) = \frac{1}{\beta n} \ ,
\eeq

\noindent where $n$ is the refractive index of the material. In
general, the refractive index varies with the photon energy: $n=n(E)$
({\it chromatic dispersion}). Since $|\cos(\Theta_C)|\leq 1$,
Cherenkov radiation is only emitted above a threshold velocity
$\beta_t=1/n$ and $\gamma_t=1/(1-\beta_t^2)^{1/2}$.

In general, Cherenkov detectors contain two main elements: a radiator
through which charged particles pass (a transparent dielectric
medium) and a photon detector. The number of photoelectrons
($N_{p.e.}$) detected in a given device can be approximated
as~\cite{pdg}

\beq
  \label{eq:richnel}
  N_{p.e.} \approx N_0 z^2 L\sin^2(\Theta_C)\ ,
\eeq

\noindent where $L$ is the path length of the particles through the
radiator, $ze$ is the particle charge and $N_0$ is a quantity called the
{\it quality factor} or {\it figure of merit}. As Cherenkov radiation
is a weak source of photons, the light transmission, collection and
detection must be as efficient as possible. These parameters are
contained in $N_0$, as well as the photon collection and detection
efficiencies of the photon detector. Typical values of $N_0$ are
between 30 and 180\,cm$^{-1}$. Three different types of Cherenkov
counters can be distinguished:

\begin{enumerate}
\item {\it Threshold counters} measure the intensity of the Cherenkov 
  radiation and are used to detect particles with velocities exceeding the
  threshold $\beta_t$. A rough estimate of the particle's
  velocity above the threshold is given by the pulse height measured
  in the photon detector.
\item {\it Differential counters} focus only Cherenkov photons with a
  certain emission angle onto the detector and in this way detect
  particles in a narrow interval of velocities.
\item {\it Imaging Cherenkov detectors} make maximum use
  of the available information (Cherenkov angle and number of photons)
  and can be divided in two main categories: RICH (Ring Imaging
  CHerenkov) and DIRC (Detection of Internally Reflected Cherenkov
  light) devices.
\end{enumerate}

In this paper the focus lies on RICH devices, since the detectors used
at the LHC experiments and discussed here all fall into this category.

\subsection{Cherenkov ring imaging}
\label{sec:richmeth}

RICH detectors resolve the ring shaped image of the focused Cherenkov
radiation. From the knowledge of the particle momentum $p$ and
Cherenkov angle $\Theta_C$ a determination of the mass of the charged
particle is possible. Combining Eqs.~\ref{eq:tof1} and \ref{eq:cangle}
yields

\beq
  \label{eq:ch}
  m = \frac{p}{c} \sqrt{n^2\cos^2(\Theta_C)-1} \ .
\eeq

\noindent In a RICH device, the Cherenkov radiation is emitted in the
radiator and collected by a photon detector, usually after being
transmitted by optical means. The first RICH was developed by
A.~Roberts in 1960~\cite{roberts}. The particular design was limited
by small angular acceptance and surface area and by low quantum and
single electron counting efficiencies. These problems were overcome by
J.~Seguinot et al. in 1977~\cite{seguinot}: A wire chamber was
converted into an efficient single photon detector by adding a
photosensitive molecule into the gas mixture and replacing one of
the cathode planes by a wire mesh and a UV transparent window.

\subsubsection{Radiator}

A RICH detector is designed to measure velocities in a specified
momentum range by using a Cherenkov radiator with refractive index
$n$ chosen such that the Cherenkov angle varies with velocity,
from threshold to the highest anticipated momentum. The thickness $L$
of the radiator is adjusted in order to assure a sufficient number of
photoelectrons for the given momentum range (see
Eq.~\ref{eq:richnel}). Two common RICH designs can be distinguished,
according to the value of $n$:

\begin{enumerate}
\item If a dense medium (large $n$) is used, only a thin radiator
  layer ($\sim$1\,cm) is required to emit a sufficient number
  of Cherenkov photons. The photon detector is located some
  distance away behind the radiator ({\it expansion gap}, usually
  about 5 to 10\,cm), allowing the light
  cone to expand and form the characteristic ring-shaped image. A
  detector designed in such a way is called {\it proximity-focusing}
  (i.e., the focusing is achieved by limiting the emission region of
  the radiation). An example is shown in Fig.~\ref{fig:hmpid}.
\item If a gaseous medium ($n\approx 1$) is used, particles have to pass a
  thicker layer ($\gtrsim$50\,cm) in order to emit a sufficient number
  of Cherenkov photons. In general, fluorocarbon gases are chosen
  because they have a low chromatic dispersion (i.e. $n$ does not
  depend strongly on the photon energy). The light is focused by
  spherical or parabolic mirrors onto the photon detectors where
  ring-shaped images are formed.  An example is shown in
  Fig.~\ref{fig:lhcbrich}.
\end{enumerate}

\subsubsection{Optics}

The quality of the optics of a RICH system influences the precision in
the Cherenkov angle measurement and the figure of merit $N_0$. Mirrors
should have high reflectivity to avoid photon loss. Traditional
constructions use a glass substrate with coating of Al for the
reflective surface and MgF$_2$ or SiO$_2$ for protection. Such mirrors
achieve values for the reflectivity of $\sim$90\%. For applications
where a minimum of material budget is required, mirrors based on
carbon fibre or Be substrates are used (see
e.g. ~\cite{lhcbrich1mirror1} and \cite{lhcbrich1mirror2}).

\subsubsection{Photon detection}

Cherenkov photons are converted to photoelectrons in photocathodes,
typically made from CsI or bialkali with low work functions. The
photon detection efficiency for a given photon detector is the product
of the quantum efficiency (the probability that an incident 
photon produces a photoelectron, typically 20 to 30\%) and the
collection efficiency (the efficiency for detecting the photoelectron,
typically 80 to 90\%). Photon detectors for RICH systems generally
fall in two categories: wire chambers with CsI coated cathode pads or
detectors based on vacuum tubes with bialkali photocathodes.
Examples of both categories are described in the following
(Sects.~\ref{sec:alicehmpid} and \ref{sec:lhcbrich}).  Low noise is a
general requirement to ensure the detection of single photoelectrons.
Recently also the readout speed of the devices became increasingly
important in order to cope with the highest event rates.

\subsubsection{Pattern recognition}

In the busy environment of hadronic collisions (such as at the LHC)
many tracks may pass through the detector, which leads to overlapping
rings. The assignment of photon hits to rings and the association of
rings to tracks requires {\it pattern recognition}. Most approaches
rely on the use of the particle track prolonged from the tracking
system to seed the ring search. After transformation through the
optics of the RICH, the track will lie inside the ring. For RICH
designs where the track is known to always point through the ring
center the ring search is actually quite simple and corresponds to the
search for a peak in the number of photon hits versus radius from the
track. In any case, the ring search efficiency needs to be studied in
detail using Monte Carlo methods.

\subsection{Angular resolution and separation power}
\label{sec:richres}

The particle velocity $\beta$ can be calculated from the reconstructed
Cherenkov angle according to Eq.~\ref{eq:cangle}. The resolution of the
velocity measurement is given by~\cite{pdg,ypsilantis}

\beq
  \label{eq:betares}
  \frac{\sigma_\beta}{\beta} = \tan(\Theta_C)\,\sigma_{\Theta_C} \ ,
\eeq

\noindent where $\sigma_{\Theta_C}$ is the resolution of the Cherenkov
angle measurement. In practical counters $\sigma_{\Theta_C}$ varies
between about 0.1 and 5\,mrad depending on the size, radiator material
and length, and on the photon detector. The Cherenkov angle is
determined by $N_{p.e.}$ measurements of the angles of emission of the
single Cherenkov photons. With the average angular
resolution for the angle measurement for the single photoelectron
$\sigma_{\Theta_i}$, the total resolution becomes

\beq
  \label{eq:betares}
  \sigma_{\Theta_C}^2 =
  \left(\frac{\sigma_{\Theta_i}}{\sqrt{N_{p.e.}}}\right)^2
  +\sigma_{Glob}^2 \ .
\eeq

\noindent The term $\sigma_{Glob}$ combines all contributions that are
independent of the single photoelectron measurement. These include
misalignment of the photon detector modules, the resolution
deterioration due to multiple scattering and background hits and
errors in the calculation of the reconstructed track parameters.
For the single photoelectron angular resolution, the following
contributions have to be considered:
\beq
  \label{eq:richressum}
  \sigma_{\Theta_i}^2 \ = \ \sigma_{EP}^2 + \sigma_{Chro}^2
  + \sigma_{Det}^2 \ ,
\eeq

\noindent where

\begin{enumerate}
\item $\sigma_{EP}$ is the geometrical error related to the emission
  point. For proximity-focusing devices the emission point is not
  known, and all photons are in general treated as if emitted at one
  point of the track through the radiator. This leads to some smearing
  of the reconstructed angle. Also for other designs the geometry of
  the optics (e.g.~the tilting of the focusing mirror) can lead to a
  dependence of the image of a Cherenkov photon on its emission
  point on the track. 
\item $\sigma_{Chro}$ is the error due to the chromatic
  dispersion of the radiators. Cherenkov photons are emitted with a
  distribution of wavelengths. Since the refractive index varies with
  wavelength, the Cherenkov angle is also spread.
\item $\sigma_{Det}$ is the error introduced by the finite granularity
  of the detector (e.g.~pixel or pad size), which limits the
  precision.
\end{enumerate}

These contributions depend on the signal-to-noise ratio (detector
gain) and can be a function of the impact angles of the track with
respect to the detector surface.

Assuming two particles of momentum $p$ and mass $m_A$ and
$m_B$, respectively, then the Cherenkov angles $\Theta_{C,A}$ and
$\Theta_{C,B}$ are measured. The resolution $\sigma_{\Theta_C}$
determines the momentum range over which the two particles can be
separated, since the Cherenkov angle saturates and thus the ring
radii get closer. Using Eq.~\ref{eq:sep}, the separation power can be
calculated as

\beq
  \label{eq:richsep1}
  n_{\sigma_{\Theta_C}} = \frac{\Theta_{C,A} -
    \Theta_{C,B}}{\langle\sigma_{\Theta_C}\rangle} \ .
\eeq

\noindent For velocities $\beta \approx 1$ well above threshold
$\beta_t$, the separation power can be approximated as~\cite{pdg}

\beq
  \label{eq:richsep2}
  n_{\sigma_{\Theta_C}} \approx
  \frac{c^2}{2p^2\langle\sigma_{\Theta_C}\rangle\sqrt{n^2-1}}
  |m_B^2-m_A^2| \ .
\eeq

\noindent Note the similarity to the formula for the separation power
using the TOF technique (Eq.~\ref{eq:tofsep}). In the case of a RICH
there is however an additional factor of $1/\sqrt{n^2-1}$, which
allows to adjust the detector configuration in order to achieve the
desired momentum coverage. Together with Eqs.~\ref{eq:richnel} and
\ref{eq:betares} it is clear that the best separation is achievable by
designing counters able to detect the maximum number of photons (large
$N_0$) with the best single photoelectron angular resolution (small
$\sigma_{\Theta_i}$).

\subsection{ALICE HMPID detector}
\label{sec:alicehmpid}

The ALICE High-Momentum Particle IDentificaton
(HMPID)~\cite{hmpid} was designed to enhance the PID capabilities of
the ALICE experiment. The HMPID is a RICH system optimised to
extend the useful range for K/$\pi$ and K/p discrimination up to
3\,GeV/c and 5\,GeV/c, respectively. Moreover, light nuclei (alpha
particles, deuterons, tritium) can be identified. The HMPID is a
single-arm array with a reduced geometrical acceptance of $|\eta
|<0.6$ in pseudo-rapidity and 1.2$^\circ < \phi < 58.8^\circ$ in
azimuth. The detector is based on proximity-focusing RICH counters
and consists of an array of seven identical modules of about
$1.5\times 1.5$\,m$^2$ each.

\begin{figure}[t]
  \setlength{\unitlength}{1cm}
  \begin{minipage}[t]{6cm}
  \includegraphics[width=6cm]{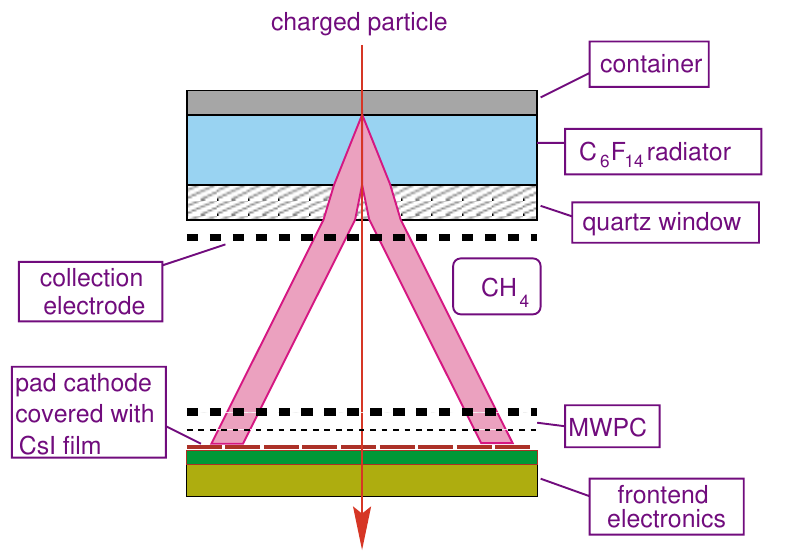}
  \end{minipage}
  \hfill
  \begin{minipage}[t]{6cm}
    \includegraphics[width=6cm]{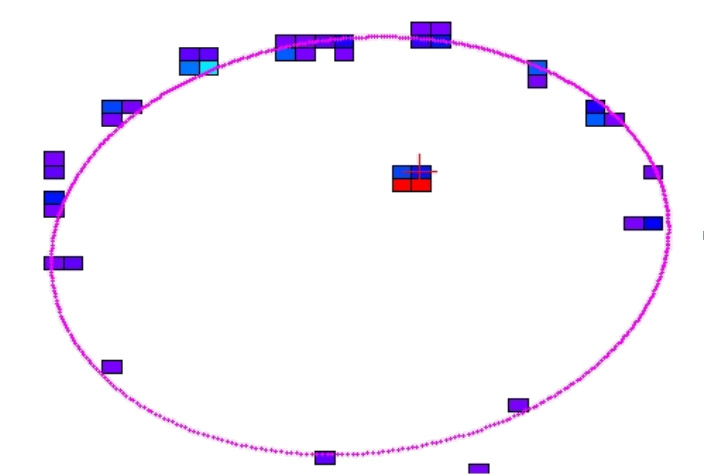}
  \end{minipage}
  \hfill
  \begin{minipage}[b]{12cm}
 \caption{Left image: schematic layout of an ALICE HMPID module,
   showing radiator, expansion gap and photon
   detector~\cite{alice}. Right image: example ring as seen by the
   HMPID. The ring has a radius of the order 10\,cm, but its shape is
   distorted due to the impact angle of the particle.}
  \label{fig:hmpid}
  \end{minipage}
\end{figure}

\subsubsection{Radiator}
 
In each module, the radiator is a 1.5\,cm thick layer of C$_6$F$_{14}$
liquid (perfluorohexane) with an index of refraction of $n\approx 1.3$
at $\lambda =175$\,nm. The material has a low chromaticity, keeping
small the corresponding term in the angular resolution. The expansion
gap has 8\,cm thickness.

\subsubsection{Photon detector}

Cherenkov photons are detected by wire chambers with a thin layer
(300\,nm) of CsI deposited onto the cathode pad plane. CsI is a stable
alkali halide crystal with high quantum efficiency of 25\% at a
wavelength of 175\,nm~\cite{csl,mauro}. The total size of the
photosensitive area is 11\,m$^2$, making the HMPID the largest scale
application of the CsI technique so far. The expansion gap and wire
chambers are filled with CH$_4$ at atmospheric pressure. The size of
the cathod pads is $8\times 8.4$\,mm$^2$.

\subsubsection{Performance and outlook}

 
\begin{figure}[t]
  \centering
  \includegraphics[width=9cm]{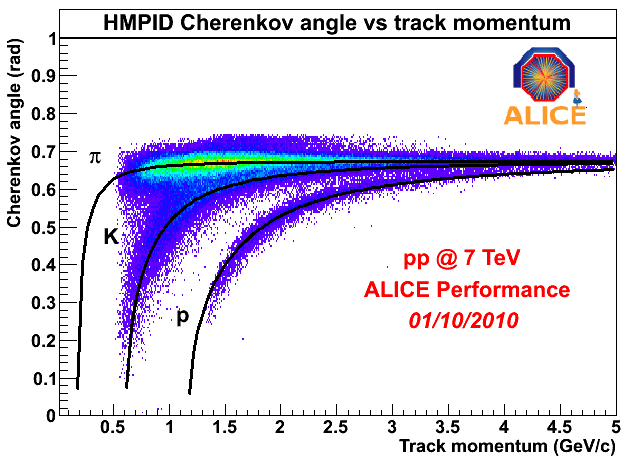}
  \caption{Dependence of the Cherenkov angle measured by the ALICE
    HMPID on the particle momentum. The lines are the theoretical
    curves calculated using Eq.~\ref{eq:cangle} with the refractive
    index $n=1.3$. No data is available for momenta below about
    600\,MeV/c, because only particles with higher momenta reach the
    detector.}
  \label{fig:hmpidrichtheta}
\end{figure}

 The identification of charged particles with known momenta in the
HMPID requires their tracks to be extrapolated from the central
tracking devices (ITS, TPC and TRD) and associated with an ionization
cluster in the HMPID. For the particular setup of a proximity-focusing
RICH, the shape of the ring is determined by the incident angle of the
particle track (see Fig.~\ref{fig:hmpid}), which makes the Cherenkov
angle reconstruction quite complex. With a single photoelectron
angular resolution $\sigma_{\Theta_i}\lesssim 12$\,mrad for perpendicular
tracks~\cite{hmpid2010}, the resolution in the reconstruction of the
Cherenkov angle is $\sigma_{\Theta_C}\approx 3.5$\,mrad (for
perpendicular tracks and for $\beta\approx 1$). Measured Cherenkov
angles for different particles as a function of momentum are shown in
Fig.~\ref{fig:hmpidrichtheta}. In the high multiplicity environment of
Pb--Pb collisions the detection of Cherenkov rings is more challenging
and the angular resolution is slightly worse. However, at the time of
writing this article no quantitative results from the first heavy-ion
data taking period (November 2010) were available to the author yet.


In order to extend the PID range up to around 30\,GeV/c, it was
proposed to add a Very High Momentum PID (VHMPID) detector to
ALICE~\cite{vhmpid}. The physics requirements have driven the
choice towards a RICH detector using a C$_4$F$_{10}$ gaseous
radiator ($n=1.0014$) with length $L=80$\,cm in a focusing
configuration using a spherical mirror.

\subsection{LHCb RICH}
\label{sec:lhcbrich}

\begin{figure}[t]
  \setlength{\unitlength}{1cm}
  \begin{minipage}[t]{5cm}
    \includegraphics[width=5cm]{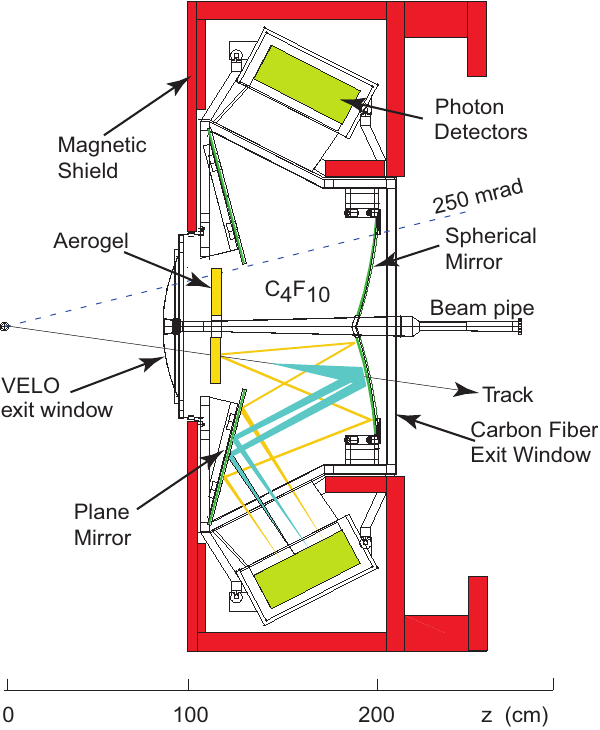}
  \end{minipage}
  \hfill
  \begin{minipage}[t]{8cm}
    \includegraphics[width=7cm]{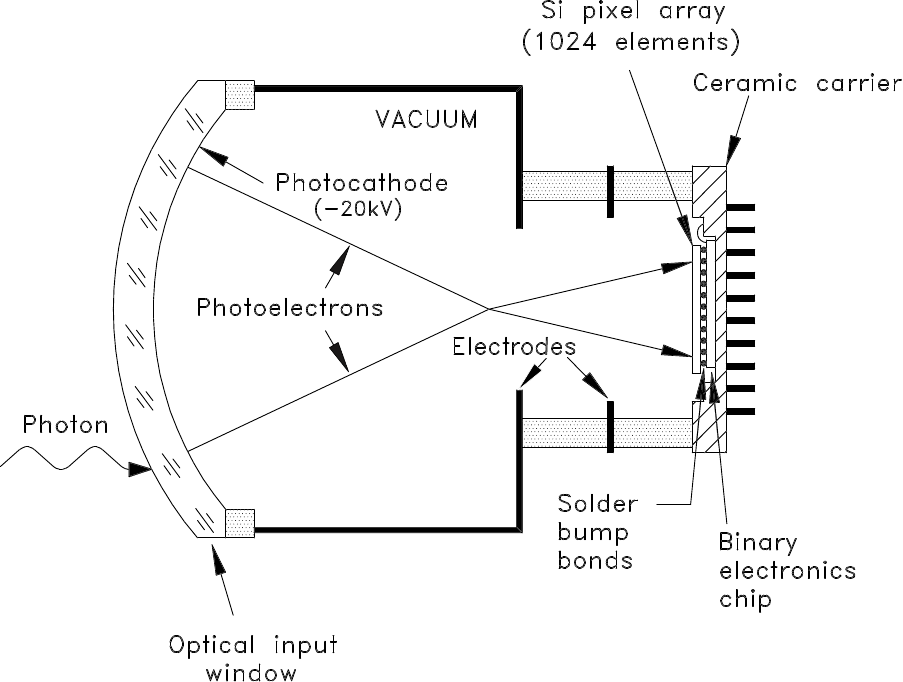}
  \end{minipage}
  \hfill
  \begin{minipage}[b]{12cm}
    \caption{Left image: schematic layout of the LHCb RICH1
      detector. The acceptance of $\pm 250$mrad is indicated. Right image: 
      schematic view of an HPD of the LHCb RICH system. Both figures
      are taken from Ref.~\cite{lhcb}.} 
   \label{fig:lhcbrich}
  \end{minipage}
\end{figure}

The LHCb experiment was introduced in Sect.~\ref{sec:lhcb}. Important
for the underlying physics goals is a K/$\pi$ separation in the
momentum range $2\lesssim p\lesssim$100\,GeV/c, which is achieved by
two RICH detectors~\cite{lhcb,lhcbrichtdr,lhcbrich2}. RICH1 and RICH2
are both characterised by a design that places the photodetectors
outside the acceptance of the LHCb spectrometer in order to limit the
degradation of the resolution of the tracking systems due to
interaction with the detector material. A set of spherical and flat
mirrors projects the
Cherenkov light onto the detector plane. The spherical mirrors are
placed inside the spectrometer acceptance and special care was taken
to minimise the fraction of radiation length while at the same time
ensuring the mechanical integrity of the optics. The spherical mirror
of RICH1 is made from carbon fibre, with thickness corresponding to
only about $0,01\,X_0$. The schematic layout of RICH1 is shown in
Fig.~\ref{fig:lhcbrich}.

\subsubsection{Radiators}

\begin{figure}[p]
  \centering
  \includegraphics[width=9cm]{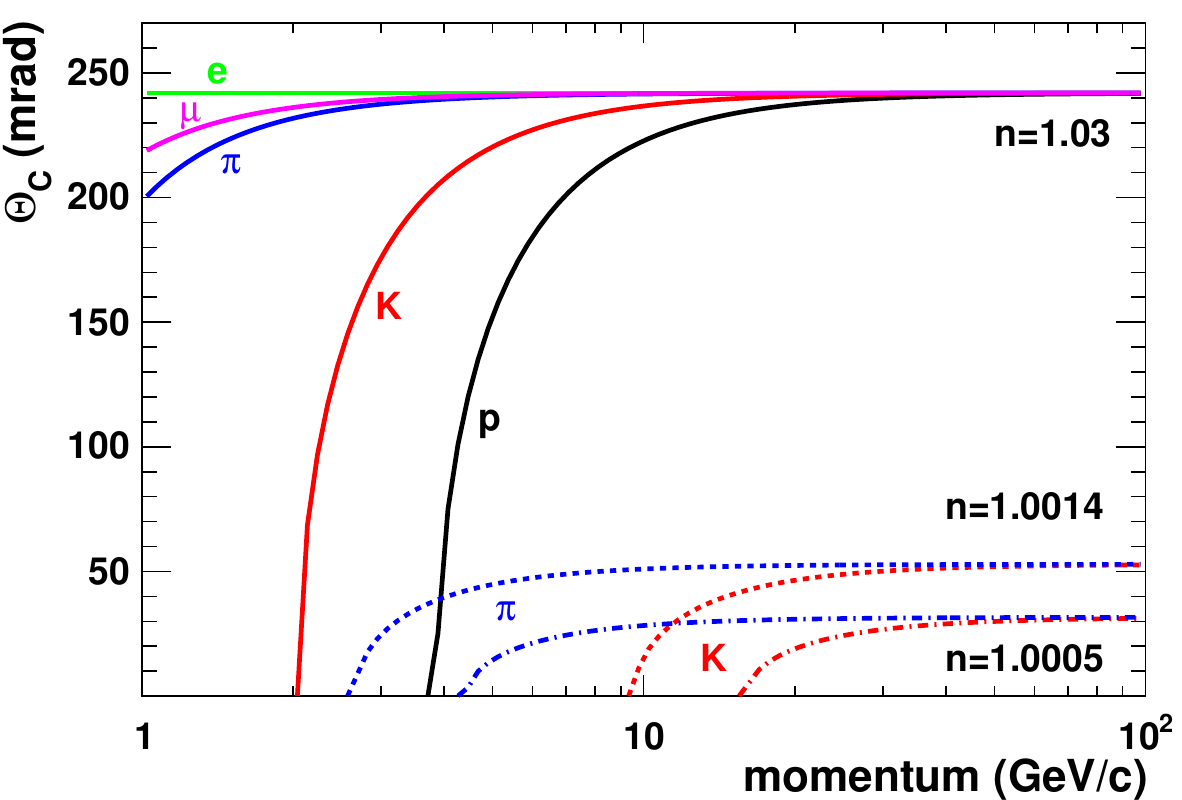}
  \caption{Cherenkov angles as a function of momentum for
    different particle species and for the three different values of
    the refractive index $n$ corresponding to the three radiator
    materials used in the LHCb RICH setup.}
  \label{fig:lhcbrichtheta}
\end{figure}

In both detectors gaseous radiators are used. In RICH1 a second,
denser radiator made from silica aerogel\footnote{Silica aerogel
  provides refractive indices that range between about 1.0026 and
  1.26~\cite{aero1,aero2}, closing the gap between gases
  ($n\approx 1$) and liquids/solids ($n\gtrsim 1.3$).}  is installed
in addition. Together the three radiators cover the large momentum
range that is required. The parameters of the three radiators are
summarised in Tab.~\ref{tab:lhcbrich}. The Cherenkov angles for the
three different materials and for different particles as a function of
the momentum are shown in Fig.~\ref{fig:lhcbrichtheta}.

\begin{table}[ht]\footnotesize
  \caption{Some parameters of the LHCb RICH detectors. The quoted
    measured angular resolutions~\cite{blanks} are for the preliminary
    alignment available from the first data sample with p--p
    collisions at $\sqrt{s}=7$\,TeV.}
  \label{tab:lhcbrich}
  \begin{center}
    \begin{tabular}{|l|c||c|c|c|c|}
      \hline
      \multicolumn{2}{|c||} \quad & \multicolumn{2}{|c|} {RICH1} &
      RICH2 \\
      \hline
      \multicolumn{2}{|c||} \quad & Silica aerogel & C$_4$F$_{10}$ &
      CF$_4$ \\
      \hline
      \hline
      \multicolumn{2}{|l||} {Momentum range [GeV/c]} & $\leq$10 &
      $10\lesssim p\lesssim 60$ & $16\lesssim p\lesssim$100 \\
      \hline
      Angular acceptance [mrad] & vertical & \multicolumn{2}{|c|}
      {$\pm 25$ to $\pm 250$} & $\pm 15$ to $\pm 100$\\
      \cline{2-5}
      & horizontal & \multicolumn{2}{|c|} {$\pm25$ to $\pm 300$} &
      $\pm 15$ to $\pm 120$ \\
      \hline
      \multicolumn{2}{|l||} {Radiator length [cm]} & 5 & 95 & 180 \\
      \hline
      \multicolumn{2}{|l||} {Refractive index $n$} & 1.03 (1.037) &
      1.0014 & 1.0005 \\
      \hline
      \multicolumn{2}{|l||} {Maximum Cherenkov angle [mrad]} & 242
      (268) & 53 & 32\\
      \hline
      \multicolumn{2}{|l||} {Expected photon yield at $\beta\approx
        1$} & 6.7 & 30.3 & 21.9 \\
      \hline
      $\sigma_{\Theta_C}$ [mrad] & expected & 2.6 & 1.57 & 0.67 \\
      \cline{2-5}
      & measured & $\sim$7.5 & 2.18 & 0.91 \\
     \hline
 \end{tabular}
  \end{center}
\end{table}

\subsubsection{Photon detectors}
 
For the photon detectors the requirements are very demanding: they
should be sensitive to single photons with high efficiency, cover
a large area (about 4\,m$^2$) with a granularity of about
$2.5\times 2.5$\,mm$^2$, and have readout fast enough to match the
LHC bunch-crossing separation of 25\,ns. LHCb chose Hybrid Photon
Detectors (HPD)~\cite{lhcbhpd}, which combine the photon sensitivity of
vacuum PMTs with the excellent spatial and energy resolutions of
silicon sensors. The conversion of the photoelectrons takes place in
the quartz window with bialkali photocathode and they are then
accelerated through a potential difference of about 20\,kV and
detected by a silicon pixel chip. HPDs are sensitive to single photons
in the wavelength range from 200 to 600\,nm with a quantum efficiency
of about 31\%. About 65\% of the detection plane is covered with read
out pixels.

\begin{figure}[p]
  \centering
  \includegraphics[width=9cm]{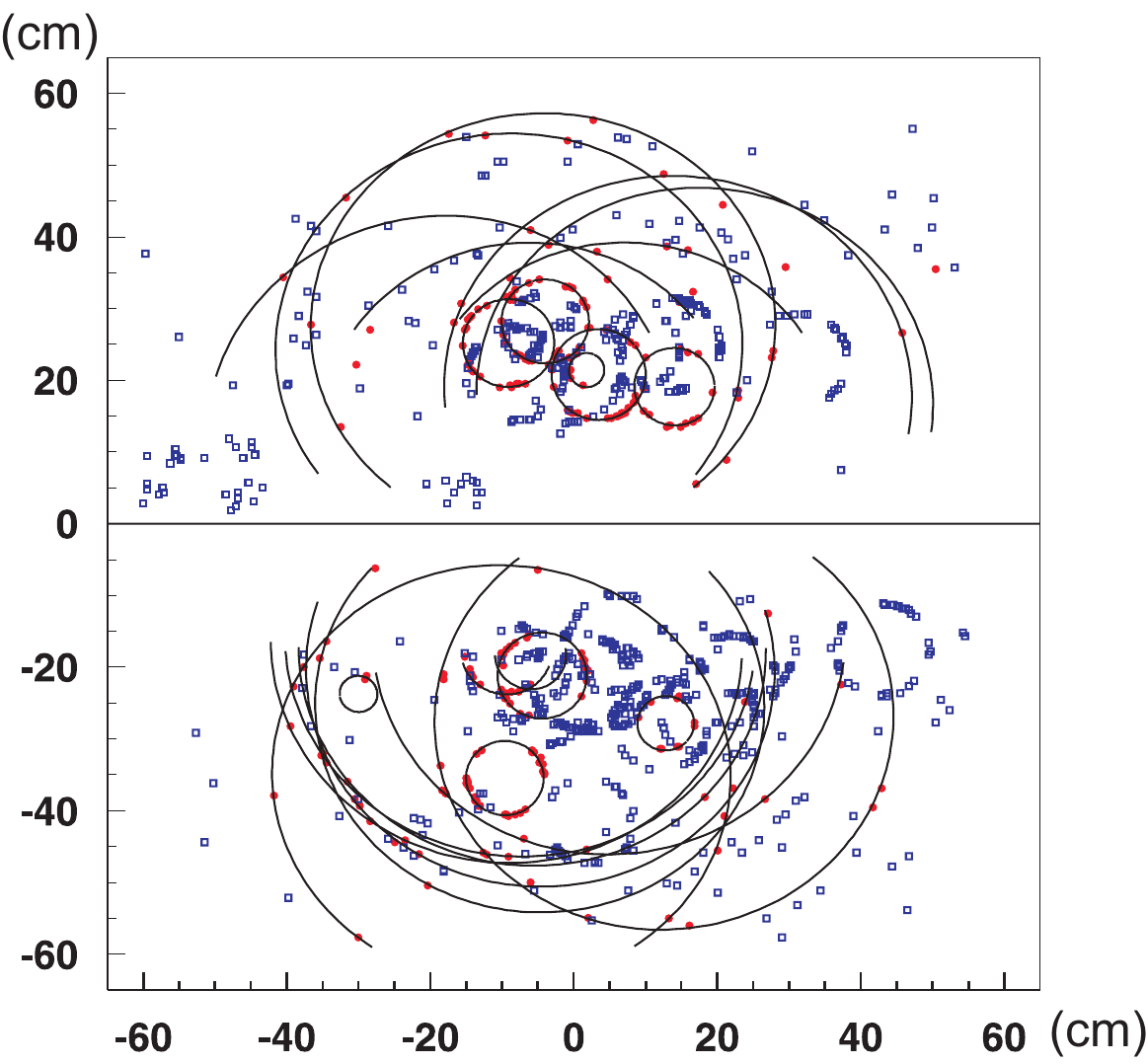}
  \caption{Typical, simulated LHCb event in the RICH1
    detector~\cite{lhcb}. The data from the two photodetector planes
    are drawn in the upper and lower halves.}
  \label{fig:richrings}
\end{figure}
 
The photoelectron trajectories are deformed by the magnetic field of
the large LHCb dipole magnet. This effect can be corrected by
comparing reproducible light-spot patterns, that are illuminated onto
the photon detector plane, with magnetic field on and off.

\subsubsection{Performance}

\begin{figure}[t]
  \centering
  \includegraphics[width=9cm]{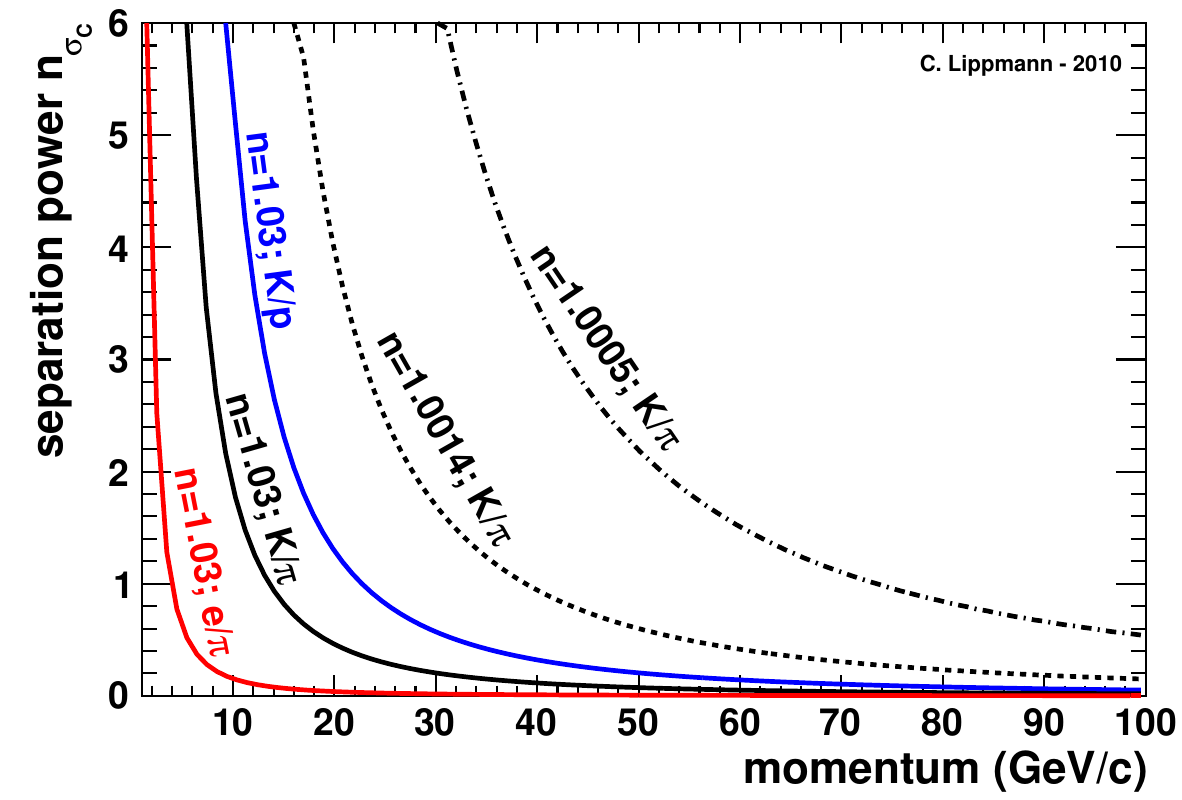}
  \caption{Particle separation achievable with Cherenkov angle
    measurements for the three different radiator materials used in the
    LHCb RICH detectors. As Cherenkov angular resolutions the expected
    values from Tab.~\ref{tab:lhcbrich} were used.}
  \label{fig:richsep}
\end{figure}

A typical (simulated) LHCb event in the RICH1 detector is shown in
Fig.~\ref{fig:richrings}. The expected single photoelectron angular
resolutions are given in Tab.~\ref{tab:lhcbrich}. After the first
alignment carried out with data from p--p collisions, the
achieved resolutions are already approaching those
values~\cite{blanks}. For the aerogel radiator, the refractive index
was found to change from its initial value of $n=1.03$ to $n=1.037$ in
the experiment, which is probably explained by absorption of
C$_4$F$_{10}$, since the aerogel material is not sealed from the gas
volume. As a consequence, the measured photon yield and angular
resolution are slightly worse than expected. Using the expected
values, the separation power $n_{\sigma_{\Theta_C}}$ can be calculated for
all three radiator materials according to Eq.~\ref{eq:richsep1}. The
results are shown in Fig.~\ref{fig:richsep}.

Particle identification with the LHCb RICH system uses a global
pattern recognition and maximum likelihood method. All found
tracks in the event and all three radiators are considered
simultaneously and the observed pattern of hit pixels in the RICH
photodetectors is matched to that expected from the reconstructed
tracks under a given set of particle hypotheses. The likelihood is
maximised by varying the particle hypotheses of each track to be an
electron, muon, pion, kaon or proton. In this way efficient K/$\pi$
separation is achieved with high purity~\cite{lhcb}: the average
efficiency for kaon ID for momenta between 2 and
100\,GeV/c is at the level of 90\%, with a corresponding average pion
misidentification rate below 5\%.

\subsection{Other Cherenkov detectors}
\label{sec:richother}

Cherenkov counters have found applications in many particle physics
experiments in many different configurations, as is documented in the
proceedings of the RICH workshop
series~\cite{rich1,rich2,rich3,rich4,rich5,rich6,rich7}. To
demonstrate this variety, here a selection of other Cherenkov
detectors ouside the LHC is given.

\begin{itemize}
\item A RICH with a 16\,m long He ($n=0.000035$) radiator,
  8\,m focal length mirrors and wire chambers with triethylamine (TEA)
  as photosensitive component was used in a spectrometer at
  FNAL~\cite{fnal1,fnal2} and provided K/$\pi$ separation up to
  200\,GeV/c.
\item The HERA-B collaboration showed that a RICH with gaseous
  (C$_4$F$_{10}$) radiator 
  and multianode PMT readout can be safely operated at high track
  densities, and no degradation of performance was observed in five
  years of operation~\cite{herab}.
\item The central region of the RICH of the COMPASS experiment was
  upgraded by replacing the wire chamber based photon detector with
  multianode PMTs very similar to the ones employed in the HERA-B
  experiment~\cite{compass}.
\item In the Belle spectrometer a Cherenkov counter based on an
  aerogel radiator and fine mesh PMT readout is used in a
  threshold configuration~\cite{acc}: The refractive index is chosen
  such that pions emit Cherenkov radiation while kaons stay below
  threshold. 
\item At the CLEO-III experiment a proximity-focusing RICH counter
  based on LiF as solid radiator and wire chambers with TEA as
  photosensitive component is installed~\cite{cleo}.
\item The RICH of the HADES experiment employs a gaseous
  (C$_4$F$_{10}$) radiator and a gaseous photon detector with
  CsI converter and acts as a hadron blind trigger
  device~\cite{hades}.
\item The Hadron Blind Detector (HBD) of the PHENIX experiment is
  built for electron ID. It uses a gaseous radiator
  (CF$_4$) and as photon detector three layers of Gas Electron
  Multipliers (GEM) with CsI coating on the first layer~\cite{phenix}.
\item The DIRC of the BaBar experiment~\cite{babar} uses 4.9\,m long,
  rectangular bars made from synthetic fused silica (average
  refractive index $n\approx 1.473$) as Cherenkov radiator and light
  guide. The photons are focussed by a ``pin-hole'' and the image is
  expanded through a standoff region filled with 6000\,l of purified
  water onto an array of PMTs placed at a distance of about 1.2\,m
  from the bar end.
\end{itemize}

\subsection{Developments for future RICH detectors}
\label{sec:richnew}

The Cherenkov counters of the BaBar and Belle spectrometers have
contributed significantly to the recent advances in flavor
physics. In both cases, upgrades are planned to further improve the
performance and to allow them to work at higher event and
background rates. The LHCb collaboration is planning to
upgrade the detector around the year 2016, following first data taking
phase~\cite{lhcbup}. The upgrade strategy involves increasing the
design luminosity by a factor of 10 to about
$2\times 10^{33}$\,cm$^{-2}$s$^{-1}$. In that case the full experiment
is read out at 40\,MHz (currently 1\,MHz). As a consequence the
HPDs need to be replaced with a faster technology. There are a few
promising candidates, however, here the developments in the wide area
of photon detectors are not discussed. A summary is given for example
in Refs.~\cite{photondet,silicondet}.

\begin{figure}[t]
  \centering
  \includegraphics[width=6cm]{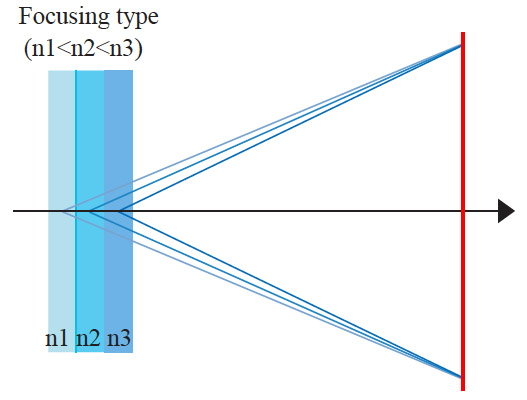}
  \caption{Schematic view of a proximity-focusing RICH with an
    inhomogeneous aerogel radiator in the focusing configuration.
    Cherenkov photons from the different radiator layers overlap, thus
    minimising the error in the angular resolution due to the
    uncertainly in the emission point~\cite{aero3}.}
  \label{fig:aero}
\end{figure}

\subsubsection{Proximity-focusing aerogel RICH}

The drawback of the proximity-focusing RICH technology is that for
thicker radiators (needed to increase the number of photons) the
emission point uncertainty ($\sigma_{EP}$ in Eq.~\ref{eq:richressum})
leads to a degradation of the single photon resolution. This
limitation can be overcome by using a radiator made from several
layers of aerogel with refractive indices gradually increasing, as is
shown in Fig.~\ref{fig:aero}. The corresponding Cherenkov rings are
overlapping on the photon detector, which leads to an optimised
angular resolution. Such a proximity-focusing RICH with inhomogeneous
aerogel radiator is proposed for the endcap region of the upgraded
Belle spectrometer~\cite{aero4,aero5}.

\begin{figure}[t]
  \centering
  \includegraphics[width=9cm]{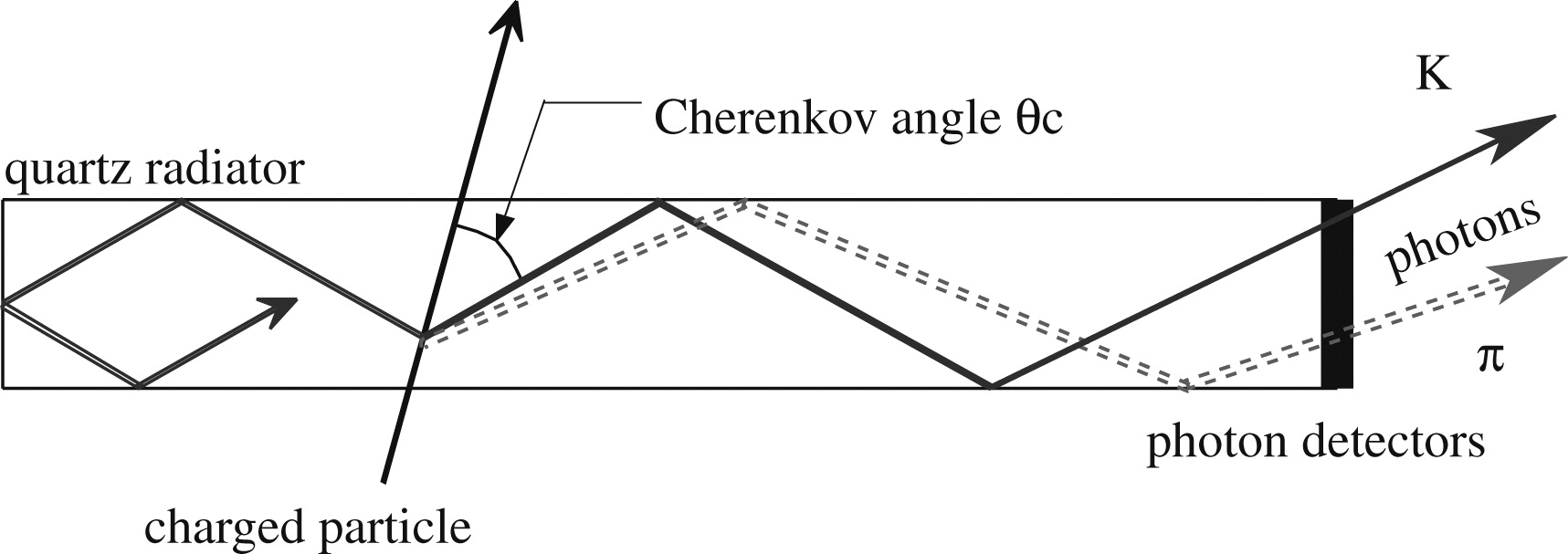}
  \caption{Schematic side view of a TOP counter~\cite{top1}. Cherenkov
    photons are guided to the photon detector by total internal
    reflections. The difference in the propagation time for two
    particle types (here kaons and pions) is due to the emission angle
    of the Cherenkov radiation and can be used to enhance PID
    information.}
  \label{fig:top}
\end{figure}

\subsubsection{Time-of-propagation counters}

For the barrel region of the upgraded Belle spectrometer, a
Time-of-propagation (TOP) counter is being
studied~\cite{top1,top2}. Here the two dimensional information of a
Cherenkov ring image is represented by the time-of-arrival and impact
position of the photons at the quartz bar exit window of a DIRC (see
Fig.\ref{fig:top}). The construction is more compact than a DIRC,
since the large expansion volume is not needed. A similar concept
combining TOF and RICH techniques, named TORCH, was proposed for the
LHCb upgrade as a possible solution for the identification of low
momentum hadrons ($p<10$\,GeV/c)~\cite{torch}. It uses a large quartz
plate to produce Cherenkov light and identifies the particles by
measuring the photon arrival times.

\section{PID in space}
\label{sec:nonacc}

\begin{figure}[t]
  \centering
  \includegraphics[width=8 cm]{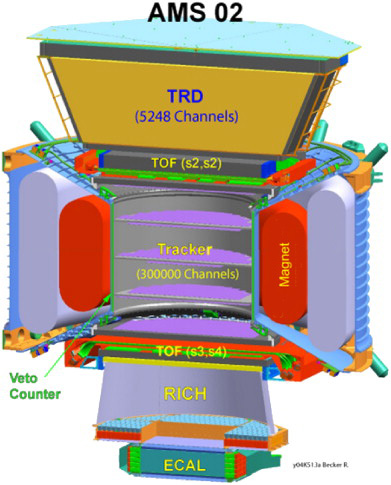}
  \caption{Schematic view of the AMS\,02 spectrometer~\cite{ams}.}
  \label{fig:ams}
\end{figure}
 
Non-accelerator experiments have become increasingly important in
particle physics. An impressive example of a space-based experiment
using a number of familiar PID techniques is the Alpha Magnetic
Spectrometer (AMS\,02~\cite{ams}, see Fig.~\ref{fig:ams}). The purposes
of the experiment are to search for cosmic antimatter and dark matter
and to study the composition and energy spectrum of the primary
cosmic rays. The AMS\,02 is set for installation on the International
Space Station (ISS). It will be transported in the cargo bay of the
space shuttle on its final mission, scheduled for 2011. The
experiment will be attached to the outside of the space station and
gather data that will be sent to a ground station for analysis.

The AMS\,02 experiment is a state-of-the-art particle physics detector
in space. The initial design included a superconducting magnet (0.9\,T),
which was however recently replaced by a permanent magnet (0.14\,T)
which allows the experiment to remain operational for the entire
lifetime of  the ISS (up to 18 years). The setup further includes a
high-precision, double sided silicon strip detector and 4 dedicated
PID devices: a TRD for electron ID (straw tubes filled with a mixture
of Xe and CO$_2$ and fleece radiators), a proximity-focusing RICH with
an aerogel radiator ($n=1.03$) for determination of the particle
velocities and charges\footnote{In the RICH the charge of the incident
  particle is found by measuring the number of  photoelectrons using
  Eq.~\ref{eq:richnel}.}, a TOF detector (four planes of plastic
scintillators) providing velocity determination and a fast trigger,
and a sampling EM calorimeter (lead absorber plates and scintillating
fibres). The whole system provides identification of particles and
isotopes up to iron.

\section{Summary and outlook}
\label{sec:sum}
 
Particle identification (PID) is an important ingredient to
particle physics experiments. Short-lived particles are reconstructed
from their decay vertex and/or decay products. Some long-lived
particles (leptons and photons) can be identified from the signatures they
leave in the different layers of a typical experiment. Distinguishing
the different long-lived charged hadrons (pions, kaons and protons) is
more challenging, but often their identification is crucial,~e.g. for
precision measurements of rare processes with high statistics. Usually
dedicated detectors are required for the task, which are based on the
determination of the particle's mass by simultaneous measurements of
momentum $p$ and velocity $\beta=v/c$. The velocity is obtained by
one of four techniques: measurement of the energy deposited through
{\it ionization}, measurement of the {\it time-of-flight}, imaging of
{\it Cherenkov radiation} or detection of {\it transition radiation}
(transition radiation is not covered by this review). An illustrative
comparison of the methods is shown in Fig.~\ref{fig:summary}. The
Cherenkov imaging method is the most flexible, as it allows to tune
the response of the detector by varying the refractive index (and the
length of the radiator), and makes accessible also very high momenta
($>$100\,GeV/c). The Cherenkov and time-of-flight techniques can be
used to close the gap in momentum in the MIP region, where PID by
ionization measurements is not possible. All techniques are very well
understood, with detectors being of the second, third or fourth
generation, but the use and sophistication is still increasing as new
generation detectors are built.

At the LHC, PID is one of the main challenges. Different concepts
are followed: the two specialised experiments, ALICE for heavy-ion
collisions and LHCb for B physics, make use of specific techniques
adapted to their physics program. In both cases PID at low transverse
momentum (but high total momentum in the case of LHCb) is
important. ALICE actually uses all existing PID techniques. The
general purpose detectors on the other hand, ATLAS and CMS, follow the
traditional setup of a particle physics experiment: a tracking system,
EM and hadron calorimeters and a muon system arranged in layers,
with large acceptance for the identification of charged hadrons,
photons and leptons. For their physics program, the challenges for
precision calorimetry and precision muon detection are
extremely high. Here the two experiments are actually following
complementary approaches, but it can be expected that they are very
competitive. All four experiments are very well advanced with the
detector calibration. They have started to record data as the LHC
began delivering p--p collisions in the year 2009. In the course of
the year 2010, the increasing luminosity allowed real stress tests of
the systems. At the end of that year also the first heavy-ion
collisions were recorded by ALICE, ATLAS and CMS.

\section*{Acknowledgements}

Thanks to D.~Di~Bari, R.~Forty, A.~Kalweit, C.~Matteuzzi,
L.~Molnar, R.~Nania, F.~Noferini, R.~Preghenella and G.~Scioli for
useful discussions and for providing figures. I am grateful to
C.~Garabatos, W.~Riegler, J.~Thomas, D.~Vranic and C.~Zampolli for
carefully reading and commenting draft versions of the paper.


\end{document}